\documentclass[fleqn,usenatbib]{mnras}

\usepackage[T1]{fontenc}

\DeclareRobustCommand{\VAN}[3]{#2}
\let\VANthebibliography\thebibliography
\def\thebibliography{\DeclareRobustCommand{\VAN}[3]{##3}\VANthebibliography}

\usepackage{graphicx}	
\usepackage{amsmath}	
\usepackage{amssymb}	

\title[Exotic arcs in RXJ0437]{RXJ0437+00: Constraining Dark Matter with Exotic Gravitational Lenses}

\author[D. J. Lagattuta et al.]
{David J. Lagattuta,$^{1,2}$\thanks{E-mail: david.j.lagattuta@durham.ac.uk}
Johan Richard,$^{3}$
Harald Ebeling,$^{4}$
Quentin Basto,$^{3}$
Catherine Cerny,$^{1,2}$
\newauthor
Alastair Edge,$^{1}$
Mathilde Jauzac,$^{1,2,5,6}$
Guillaume Mahler,$^{1,2}$ and
Richard Massey$^{1,2}$
\\
$^{1}$Centre for Extragalactic Astronomy, Department of Physics, Durham University, South Road, Durham DH1 3LE, UK\\
$^{2}$Institute for Computational Cosmology, Department of Physics, Durham University, South Road, Durham DH1 3LE, UK\\
$^{3}$Univ Lyon, Univ Lyon1, Ens de Lyon, CNRS, Centre de Recherche Astrophysique de Lyon UMR5574, 69230, Saint-Genis-Laval, France\\
$^{4}$Institute for Astronomy, University of Hawaii, 640 N Aohoku Pl, Hilo, HI 96720, USA\\
$^{5}$Astrophysics Research Centre, University of KwaZulu-Natal, Westville Campus, Durban 4041, South Africa \\
$^{6}$School of Mathematics, Statistics \& Computer Science, University of KwaZulu-Natal, Westville Campus, Durban 4041, South Africa\\
}

\date{Accepted 2023 March 10. Received 2023 March 1; in original form 2023 January 10}

\pubyear{2023}

\usepackage{newtxtext,newtxmath}
\begin{document}
\label{firstpage}
\pagerange{\pageref{firstpage}--\pageref{lastpage}}
\maketitle

\begin{abstract}

We present the first strong-gravitational-lensing analysis of the galaxy cluster RX\,J0437.1+0043 (RXJ0437; $z = 0.285$). Newly obtained, deep MUSE observations, Keck/MOSFIRE near-infrared spectroscopy, and \emph{Hubble} Space Telescope SNAPshot imaging reveal 13 multiply imaged background galaxies, three of them (at $z=1.98$, 2.97, and 6.02, respectively) in hyperbolic umbilic (H-U) lensing configurations. The H-U images are located only 20--50\,kpc from the cluster centre, i.e., at distances well inside the Einstein radius where images from other lens configurations are demagnified and often unobservable. Extremely rare (only one H-U lens was known previously) these systems are able to constrain the inner slope of the mass distribution -- and unlike radial arcs, the presence of H-U configurations is not biased towards shallow cores. The galaxies lensed by RXJ0437 are magnified by factors ranging from 30 to 300 and (in the case of H-U systems) stretched nearly isotropically. Taking advantage of this extreme magnification, we demonstrate how the source galaxies in H-U systems can be used to probe for small-scale ($\sim 10^{9} M_{\odot}$) substructures, providing additional insight into the nature of dark matter. 

\end{abstract}




\section{Introduction}
\label{sec:intro}

Galaxy clusters are ideal astrophysical laboratories for studies of the properties and distribution of mass in the Universe:\ they have large physical sizes, span a wide range in environmental density, and contain significant quantities of both baryons and dark matter \citep[DM, e.g.,][]{clowe2006}. Forming at nodes of the cosmic web and growing by accretion of infalling matter and structures from attached filaments, clusters are tightly connected to the evolution of large-scale structure in the Universe. Understanding the distribution of mass within clusters therefore provides significant insight into the cosmological model \citep[e.g.,][]{kafer2019}. While there are many ways of probing mass in clusters, gravitational lensing is an especially potent tool, since lensing-based measurements are sensitive to both baryonic and dark matter and thus yield total-mass estimates without the need for simplifying assumptions about the dynamical state of the cluster or its morphology and geometry. In particular, lensing acts as a direct, geometric probe of the mass distribution in the central cores of clusters (the area known as the ``strong-lensing regime'') where high mass densities magnify and distort the light of distant background galaxies into giant arcs and multiple-image systems. 

Since strong lensing signals can be especially sensitive to medium- and small-scale mass fluctuations, deep lensing studies have become a common way to probe the central structure of clusters \citep[e.g.,][]{grillo2015,jauzac2016,lagattuta2017,lagattuta2019,mahler2018,ghosh2021,acebron2022}. Yet, in spite of its significant advantages, the method has limitations:\ while strong lensing can reveal mass distributions over a variety of physical scales, its power is diminished in the innermost regions of clusters ($< 50$\,kpc). This is because the images of ``traditional'' lensing configurations (such as doubles, quadruples, and ring-like systems) that form in these regions are often highly de-magnified, making them extremely difficult to detect and characterize. 

The slope of the central mass distribution in clusters is an important parameter in many areas of astrophysics -- from structure formation, through galaxy evolution, to the nature of DM itself \citep[e.g.,][]{robertson2019} -- and its exact form is still a topic of considerable debate (one example being the upscaled ``cusp-core problem'' in massive ellipticals; \citealt{andrade2019}). There are, however, certain lensing configurations that, although very uncommon, allow us to probe the mass profile in the very core of clusters and measure this important physical parameter.

Specifically, ``exotic'' lens systems \citep[e.g.,][]{orban2009,meena2020}, produce images with unusually high magnification factors ($\mu=100$ or more) that are, crucially, located within the innermost $\sim$ 50\,kpc of the cluster center. Of these exotic lens configurations, Hyperbolic-Umbilic (H-U) patterns are particularly valuable, since they create the largest number of multiple images and thus provide the most robust constraints on the local mass distribution. Formally, H-U configurations occur when a cusp point in the source plane is ``exchanged'' between a radial and tangential caustic curve (see e.g. \citealt{schneider1992} or \citealt{petters01} for a mathematical definition). Practically, this means that H-U images will form in places where cluster-scale tangential and radial critical curves (the lens-plane equivalent of caustics) come very close together or even touch, naturally leading them to appear near the cluster centre. Thus, measurements derived within the H-U region serve as important anchor points for the inner mass slope, which can be compared to predictions made by the prevailing $\Lambda$CDM cosmological model \citep{harvey2019,robertson2019}. Discrepancies between theory and observation can then be used to test alternatives to $\Lambda$CDM, such as self-interacting-DM (SIDM) varieties. Historically, H-U systems have been exceedingly rare, and until recently only a single example (Abell\,1703; \citealt{limousin2008}) could be found in the literature. Fortunately, this situation is changing, and the population of exotic cluster lenses is slowly increasing, largely thanks to data obtained with the Multi-Unit Spectroscopic Explorer (MUSE; \citealt{bacon2010}) on the Very Large Telescope (VLT). This is because, as an integral field unit (IFU) spectrograph, MUSE is sensitive to emission features that are often extremely faint in broadband imaging and/or significantly contaminated by brighter cluster members, increasing the density of lensed objects that can be detected in a given field. 
 
In this work, we present results of our strong-lensing analysis of RX\,J0437.1+0043 (RXJ0437; $z=0.285$), an X-ray luminous cluster originally discovered in ROSAT All-Sky Survey data \citep{ebeling2000}. Using a powerful combination of imaging and spectroscopic data, we identify \emph{three} new H-U lens systems in the cluster, allowing us to investigate the core mass distribution in considerable detail. Our work is organized as follows:\ We summarize the observations used in our analysis in Section~\ref{sec:data}, present our redshift catalog -- which we use to derive initial physical parameters of the cluster -- in Section~\ref{sec:redshifts}, describe our approach to mass modeling in Section~\ref{sec:method}, and provide an overview of all lensing constraints in Section~\ref{sec:multiImage}. The resulting lens model is presented in Section~\ref{sec:model}. We derive resolved properties of the background galaxies observed in H-U configurations in Section~\ref{sec:source} and discuss the relevance of our findings for our understanding of dark matter in Section~\ref{sec:DMscience}, before presenting conclusions in Section~\ref{sec:conclusions}. 

Throughout this work we assume a flat cosmological model with $\Omega_{\rm \Lambda} = 0.7$, $\Omega_{\rm M} = 0.3$, and $H_{0} = 70$ km s$^{-1}$ kpc$^{-1}$. Assuming these parameters, 1 arcsecond spans 4.297 kpc at the systemic cluster redshift ($z = 0.285$). Unless otherwise specified, all magnitudes are presented in the AB system \citep{oke74}.   

\section{Data}
\label{sec:data}

\begin{figure*}
    \centering
    \includegraphics[width=\textwidth]{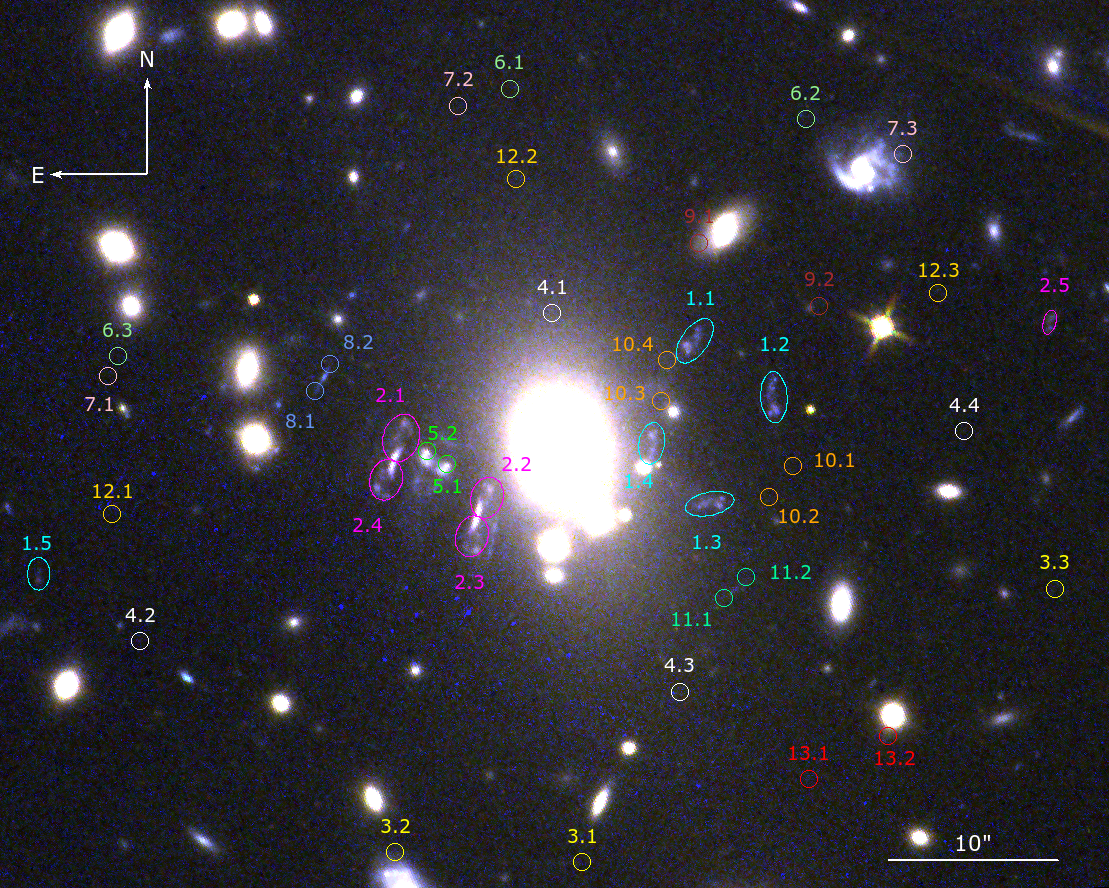}
    \caption{High-resolution colour image of the RXJ0437 cluster core obtained with the \emph{Hubble} Space Telescope (F814W/F110W/F140W). Spectroscopically confirmed multiple-image systems are overlaid in coloured ellipses. Systems 1 (cyan), 2 (magenta), and 10 (orange) are arranged in rare Hyperbolic-Umbilic (H-U) configurations.}
    \label{fig:data}
\end{figure*}

Our strong-lensing analysis of RXJ0437 is based on a combination of imaging and spectroscopic data sets. While much of our analysis focuses on MUSE and \emph{Hubble} Space Telescope (\emph{HST}) data (Fig.~\ref{fig:data}), we also derive ancillary colour information from ground-based imaging taken with the Dark Energy Camera (DECam) on the Victor M.\ Blanco Telescope at the Cerro Tololo Inter-American Observatory (CTIO).

\subsection{MUSE}
\label{sec:data_muse}

The MUSE data used in our analysis cover a $1.5\arcmin \times 1.5\arcmin$ region of the sky centred on the brightest cluster galaxy (BCG) of RXJ0437 at ($\alpha$ = 69.289677, $\delta$ = 0.73114470), spanning the cluster's strong-lensing zone and its immediate surroundings. A mosaic of nine partially overlapping 1-hour MUSE pointings forms the full field of view, which has a combined exposure-time depth ranging from 6.5 hours in the centre (near the BCG and the H-U lens systems), to 1-2 hours in the outer regions (depending on position and orientation). An initial central pointing (total exposure time 2910s) was observed on 15 February 2020 as part of the ESO Kaleidoscope Clusters survey (PID 0104.A-0801; PI A.~Edge), a large ``filler'' survey program designed to quickly identify bright strong-lensing features in shallow (snapshot) exposures of galaxy clusters.  Throughout this survey, MUSE operated in its WFM-NOAO-N mode characterized by wide-field, nominal wavelength coverage (wavelength range $4750 - 9350$ \AA, mean resolution $R = 3000$) without adaptive optics correction. The data were also obtained during grey time. Examining the data from this initial observation we detected the features of the primary H-U ring (System 1 in Section~\ref{sec:multiImage}), along with four other multiply imaged galaxies, allowing us to create a preliminary lens model and estimate the extent of the multiple-image region. Using this preliminary model as a guide, we obtained the additional data to create our final mosaic (which covers the full multiple-image footprint) in ESO program 106.21AD (PI D.~Lagattuta) between 15 January and 13 February 2021 (total exposure time ranging from 2544s to 20352s). Unlike in the initial observation, however, we did apply AO corrections (WFM-AO-N mode) during this program, and the data were obtained with stricter observational constraints: exposures were acquired in dark time (moon brightness $<$ 40\% of maximum), under clear skies ($<10\%$ cloud-cover), with natural seeing limited to $<$ 0\farcs8.

\begin{figure}
    \centering
    \includegraphics[width=0.49\textwidth,trim={40 0 20 10},clip]{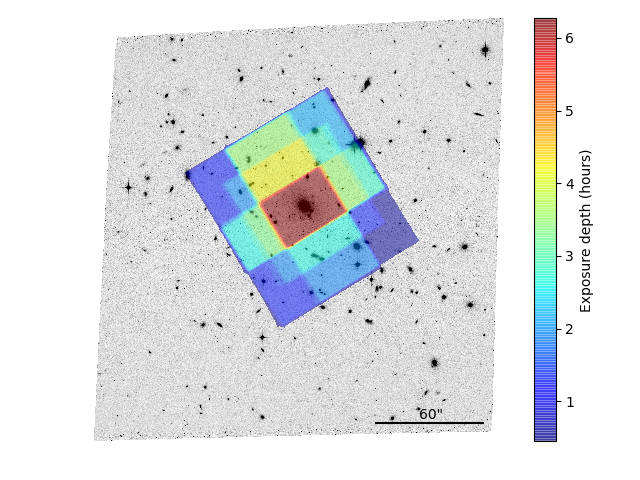}
    \caption{Footprint of the largest \emph{HST} image (F814W; grayscale) and its relative size compared to the MUSE data (coloured overlay). The colours in this overlay show the exposure time depth map of the full MUSE mosaic.}
    \label{fig:footprint}
\end{figure}

All pointings -- both in the filler survey and in dedicated follow-up campaign -- consist of a series of three (rotated) exposures taken at 0, 90, and 180 degree roll angles, respectively; each exposure is also shifted by a small dither offset (0\farcs3) to minimize the effect of bad pixels and other systematics during data combination.  However, because of differences in overheads between NOAO and AO modes, the individual AO-corrected exposures are slightly shorter (848s vs 970s) and have a coverage gap between 5805\AA\ and 5965\AA\ due to laser contamination. We reduced all data following the procedure described in Section 2.2 of \citet{lagattuta2022} to create a final, combined, mosaic data cube. The resulting MUSE exposure map is shown in Fig.~\ref{fig:footprint}.

\subsection{Keck}
\label{sec:data_keck}

RXJ0437 was observed with the near-infrared multi-object spectrograph MOSFIRE on the Keck-I 10m-telescope on Maunakea on Jan 8, 2022. We targeted two images of the multiple-image System 1 (see Section \ref{sec:multiImage}, Fig.~\ref{fig:data}, and Table~\ref{tbl:Multi-Images}
for details) for 24 min in the K band (1.92--2.40$\mu$m); a 5\arcsec\ offset was used between observations in an ABBA pattern to facilitate background subtraction. Similarly, Systems 2 and 5 were observed for 72 and 24 min in the K and H band (1.46--1.81$\mu$m), respectively, using a 1.5\arcsec\ offset. For the 0.7\arcsec\ slit used in these observations, the spectral resolution is 4.5 and 6\AA\ in H and K, respectively. All data were reduced with the MOSFIRE data reduction pipeline designed by the MOSFIRE commissioning team and written by Nick Konidaris with extensive checking and feedback from Chuck Steidel and other MOSFIRE team members.

\subsection{HST}
\label{sec:data_hst}

\emph{HST} imaging of RXJ0437 spans three distinct broadband filters:\ one in the optical (F814W) and two in the near-IR regime (F110W and F140W). Observations were acquired between November 2021 and July 2022 as part of a dedicated SNAPshot program targeting massive lensing clusters (GO-16670; PI Ebeling). The F814W image was taken with the Advanced Camera for Surveys \citep[ACS;][]{ford2003} Wide-Field Channel (WFC). The full frame is generated from three 400s exposures arranged in an \texttt{ACS-WFC-DITHER-LINE} pattern to fill in coverage of the WFC interchip gap. Individual exposures are processed and combined using the standard ACS reduction pipeline, which eliminates systematics such as hot pixels and charge-transfer inefficency (CTI) trailing, flags and removes cosmic-ray contamination, corrects for the effects of geometric distortion, and performs astrometric alignment and registration. Conversely, the F110W and F140W data were captured with the Wide-Field Camera 3 \citep[WFC3;][]{kimble2008} IR channel. The final image in each filter is the combination of two individual 353s exposures which are offset by 5\arcsec\ in a \texttt{WFC3-IR-DITHER-BLOB} pattern to again minimize the effects of bad pixels and other systematics. The final RGB image of all three filters is shown in Fig.~\ref{fig:data}. Compared to the MUSE frame, the \emph{HST} images have a much higher angular resolution:\ we measure an average point spread function (PSF) FWHM of 0\farcs15 over all \emph{HST} bands, versus 0\farcs68 for the MUSE data. The \emph{HST} images are also wider, providing a clearer picture of cluster and background features out to larger physical radii -- the \emph{HST} footprint extends to $\sim$700 kpc from the cluster core, while the MUSE footprint reaches $\sim$280 kpc. 

\subsection{DECam}
\label{sec:data_other}

Multi-band, ground-based imaging of RXJ0437 is publicly available from the Dark Energy Camera Legacy Survey (DECaLS; NOAO program 2014B-0404, PIs D.\ Schlegel \& A.\ Dey), a subset of the larger DESI Legacy Survey\footnote{\url{https://www.legacysurvey.org/}} creating wide-area mosaics covering $\sim$14000 deg$^2$ of the sky. To maximize coverage with existing data, we extract a 4 arcmin$^2$ section of the mosaics in three optical bands (g, r, and z), each centred on the RXJ0437 BCG. Individual frames making up these cutouts were taken between October 2016 and November 2018, and the final stacked mosaic frames consist of 4, 4, and 5 exposures in the g, r, and z bands, with a total exposure time of 372s, 184s, and 513s, respectively. These ground-based images have lower angular resolution than the \emph{HST} data (with an average psf of $\sim 1\farcs2$) but provide additional colour information for galaxies over a larger area. 

\begin{figure*}
    \centering
    \includegraphics[width=0.49\textwidth]{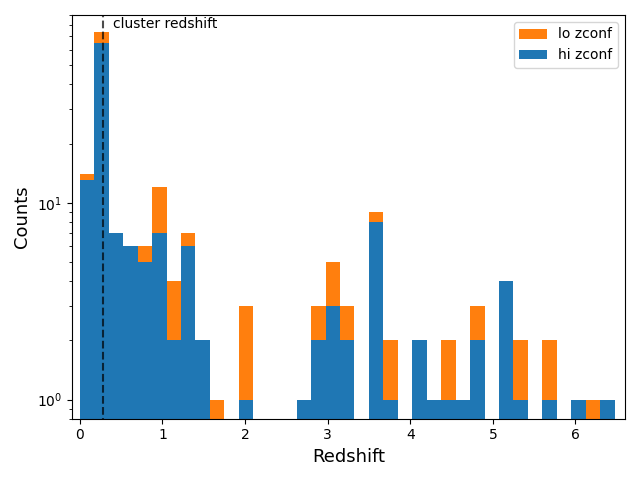}
    \includegraphics[width=0.49\textwidth]{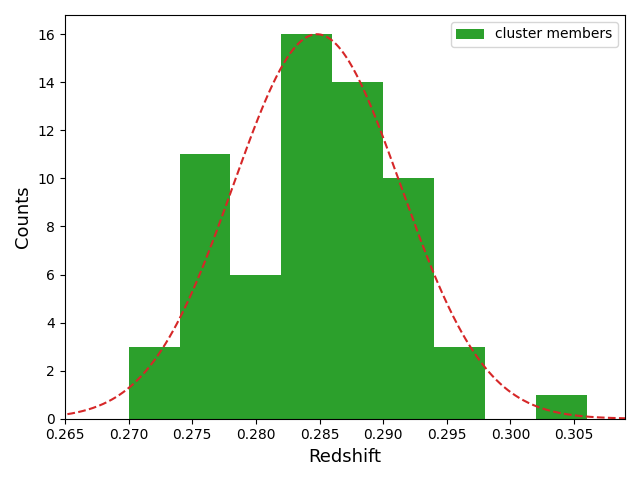}
    \caption{Left: Distribution of all galaxy redshifts measured from our MUSE data in the RXJ0437 field, after removing duplicate entries for multiply imaged galaxies. High-confidence redshifts (zconf $\geq$ 2 following e.g.\ \citealt{lagattuta2022}) are shown in blue, while lower-confidence redshifts (not used in the subsequent spectroscopic analysis) are shown in orange. The clear overdensity of detections at $z = 0.285$ represents the cluster and its systemic redshift.  Right: redshift distribution of cluster members. The 64 high-confidence redshifts shown in this histogram are well fit by a Gaussian distribution (red dashed line) characterized by a systemic redshift of 0.2847 and a velocity dispersion of 1570 km s$^{-1}$.}
    \label{fig:redshifts}
\end{figure*}

\section{Galaxy redshifts}
\label{sec:redshifts}

Prior to modelling the RXJ0437 system we measure the redshifts of objects within the MUSE footprint, in order to identify cluster members, multiple-image constraints, and look for structure along the line of sight. 

The techniques used to extract spectra and measure redshifts from the MUSE data cube are the same as those employed in previous work \citep[e.g.,][]{richard2021,lagattuta2022}, but we briefly describe them again here. We identify objects in two ways: first by selecting sources detected in the stacked \emph{HST} imaging data (which we call \emph{prior} targets), and then by scanning the muse data cube itself for prominent emission lines (\emph{muselet} targets) using the \textsc{mpdaf} software package \citep{bacon2016,piqueras2017}\footnote{\url{https://mpdaf.readthedocs.io/en/latest/muselet.html}}. Combining the sources identified by each procedure -- and matching objects to remove duplicate entries -- we then extract a spectrum for each candidate using the optimally weighted \citet{horne1986} algorithm. We use the redshift-fitting software \textsc{marz} \citep{hinton2016} to create an initial redshift guess for each object, then visually inspect the results and adjust (or reject) the values as needed. After the visual-inspection stage, we flag 171 objects as having high-confidence redshifts (measurements with a ``zconf'' value of 2 or 3, following \citealt{lagattuta2022}); this set is reduced to 147 unique objects when accounting for multiply imaged galaxies. We also measure low-confidence redshifts (zconf = 1) for 33 additional objects; although we do not include these more speculative measurements in all of the spectroscopic analysis described below, we still list them in the final redshift catalogue for completeness. In total, counting high- and low-confidence measurements and including all multiple images, we have 204 redshift entries in the RXJ0437 field.  The final catalogue is included as an electronic supplement to this work, but for clarity, we present a small subset of the entries and explain each column in Appendix \ref{sec:sampleCat}.

We present a histogram of the RXJ0437 redshift distribution -- keeping only a single entry for each multiple-image system -- in the left-hand panel of Fig.~\ref{fig:redshifts}. The full distribution extends over the redshift range $0 \leq z \leq 6.5$, with a pronounced excess of galaxies at $z \sim 0.285$ indicating the cluster redshift. We find no evidence of other major concentrations along the line of sight, with the exception of a small increase in galaxy counts at $z = 3.5$ caused by lensed background objects (almost a third of the identified multiple image systems behind RXJ0437 lie at $z \sim 3.5$; Section \ref{sec:multiImage}). Investigating the distribution further, we classify objects as foreground ($0 \leq z < 0.27$), cluster members ($0.27 \leq z \leq 0.31$), or background ($z > 0.31$), and find 16(15), 73(64), and 91(68) total(high-confidence) galaxies in each category, respectively. Focusing more closely on the cluster members (Fig.~\ref{fig:redshifts}, right) we find that the cluster redshift distribution is roughly Gaussian in shape. Converting the observed redshifts into velocity space, we measure a line-of-sight velocity dispersion of $1570^{+120}_{-160}$ km s$^{-1}$ using a biweight sample variance estimator \citep[e.g.][]{beers1990,ferragamo2020}. 

\section{Modeling technique}
\label{sec:method}

We construct a model of the mass distribution of RXJ0437 using the publicly available \textsc{Lenstool}\footnote{\url{https://projets.lam.fr/projects/lenstool/wiki}} software \citep{kneib1996,jullo2007,jullo2009}. 

The model is constructed from parametric components that represent mass halos at large and small scales. The parameters characterizing these components are constrained by the positions of multiply imaged background sources identified throughout the field.  The majority of the mass is modeled as a series of pseudo-isothermal elliptical mass distributions \citep[PIEMD;][]{eliasdottir2007}, representing both cluster- and galaxy-scale halos.  However, to account for any mass not directly observed in the available data, we also include a systematic ``external shear'' term (see e.g. \citealt{keeton1997}) as an additional component. Each PIEMD halo is described by seven parameters: spatial position ($\alpha$ and $\delta$), ellipticity and position angle ($\varepsilon$ and $\theta$), a central velocity dispersion ($\sigma_0$) normalizing the total mass, and two characteristic radii ($r_{\rm core}$ and $r_{\rm cut}$) representing respectively the inner flattening radius and outer truncation radius where the component's mass profile deviates from a purely isothermal slope. Alternatively, as a ``global'' quantity, the external shear term has only two parameters: a magnitude ($\gamma$) and position angle ($\theta_{\gamma}$) of the (spatially constant) shear. 

Having no a priori knowledge of the shape or positions of the large DM halos, we allow nearly all PIEMD parameters of cluster-scale halos to freely vary during the model optimization, although we fix the $r_{\rm cut}$ radius to a constant 800 kpc, since the typical cluster-scale truncation radius is much larger than the region covered by lensing constraints.  Conversely, we rely on observational evidence to fix several parameters of galaxy-scale halos. Specifically, we set the centroid, ellipticity, and position angle of each galaxy to values measured in the F814W image (Section~\ref{sec:data_hst}), as determined by Source Extractor \citep{bertin1996}. Additionally, we fix $r_{\rm core}$ at 0.1 kpc, based on empirical considerations of cluster galaxies \citep{limousin2007}, since again we are unable to suitably constrain this value from the data. This leaves only $\sigma_0$ and $r_{\rm cut}$ as free parameters. Rather than individually fit all galaxies, however, we instead optimize the parameters of a single characteristic galaxy (defined to be an $L^*$ galaxy at the redshift of the cluster) and rely on a mass/light scaling relation based on the Faber-Jackson relation \citep{faber1976} to generate values for all other galaxies. Thus, we need only two free parameters in total ($\sigma_0^*$ and $r_{\rm cut}^*$) to characterize the entire set of galaxy-scale potentials.

We select cluster galaxies using colour cuts designed to identify the cluster red sequence \citep{gladders2000}. Although the \emph{HST} bands are less sensitive to systematic effects, such as sky noise, and in general yield more robust photometry, we nonetheless perform the colour selection using DES data. The reasons are twofold: (1) the passbands in the DES imaging are better placed to identify the Balmer break feature critical for identifying cluster members, and (2) the DES images are slightly larger then the \emph{HST} frames, allowing us to identify potential cluster galaxies over a larger area. After identifying the phase space containing cluster members, we select a total of 143 galaxy-size halos for our model. As a further check, we inspect the redshifts of cluster member candidates that fall within the MUSE footprint, finding that all candidates in this region do indeed have spectroscopic redshifts consistent with the radial-velocity distribution of cluster members ($0.27 < z < 0.31$). We also identify $\sim$ 26 additional spectroscopically confirmed cluster members in the MUSE footprint that are not selected by the colour cut. For completeness, we include these galaxies in the lens model, but we note that they fall in the faint, blue end of the colour-magnitude diagram and should not significantly affect the mass budget. To verify this, after constructing and optimizing our primary model (Section \ref{sec:model}) we create an additional model which does not include these ``blue-end'' galaxies. Compared to the full model, the final parameters of this modified version are functionally identical, differing by less than the measurement uncertainty. In this way, we are confident that any missed blue cluster members lying outside of the MUSE footprint will have a negligible effect on the final results.   

\section{Multiple Images}
\label{sec:multiImage}

The multiple-image systems used in the model are all initially identified in the MUSE data; as a result, every multiple-image constraint has a confirmed spectroscopic redshift, providing increased precision in the final parameter values. We mark and label all multiple-image systems in a close-up view of the cluster core shown in Fig.~\ref{fig:data}; the coordinates and redshifts of all multiple-image components are listed in Table~\ref{tbl:Multi-Images}. The images 1.1, 1.3, 2.3, and 5.2 have independent spectroscopic redshifts from the Keck/MOSFIRE observations described in Section~\ref{sec:data_keck}.

As mentioned in Section~\ref{sec:data_muse}, five systems (comprising 17 individual images) were identified in the original shallow muse cube, including the bright H-U galaxy we designate as System 1. Another eight systems (consisting of 21 images) were discovered in the subsequent deeper mosaic, including an additional H-U system designated System 10. The redshifts of the identified systems range from $z = 0.9$ to $z = 6.44$, with the majority falling in the $z \sim 3 - 5$ range. Individual images are approximately evenly distributed throughout the field.

\begin{figure*}
    \centering
    \includegraphics[width=0.49\textwidth]{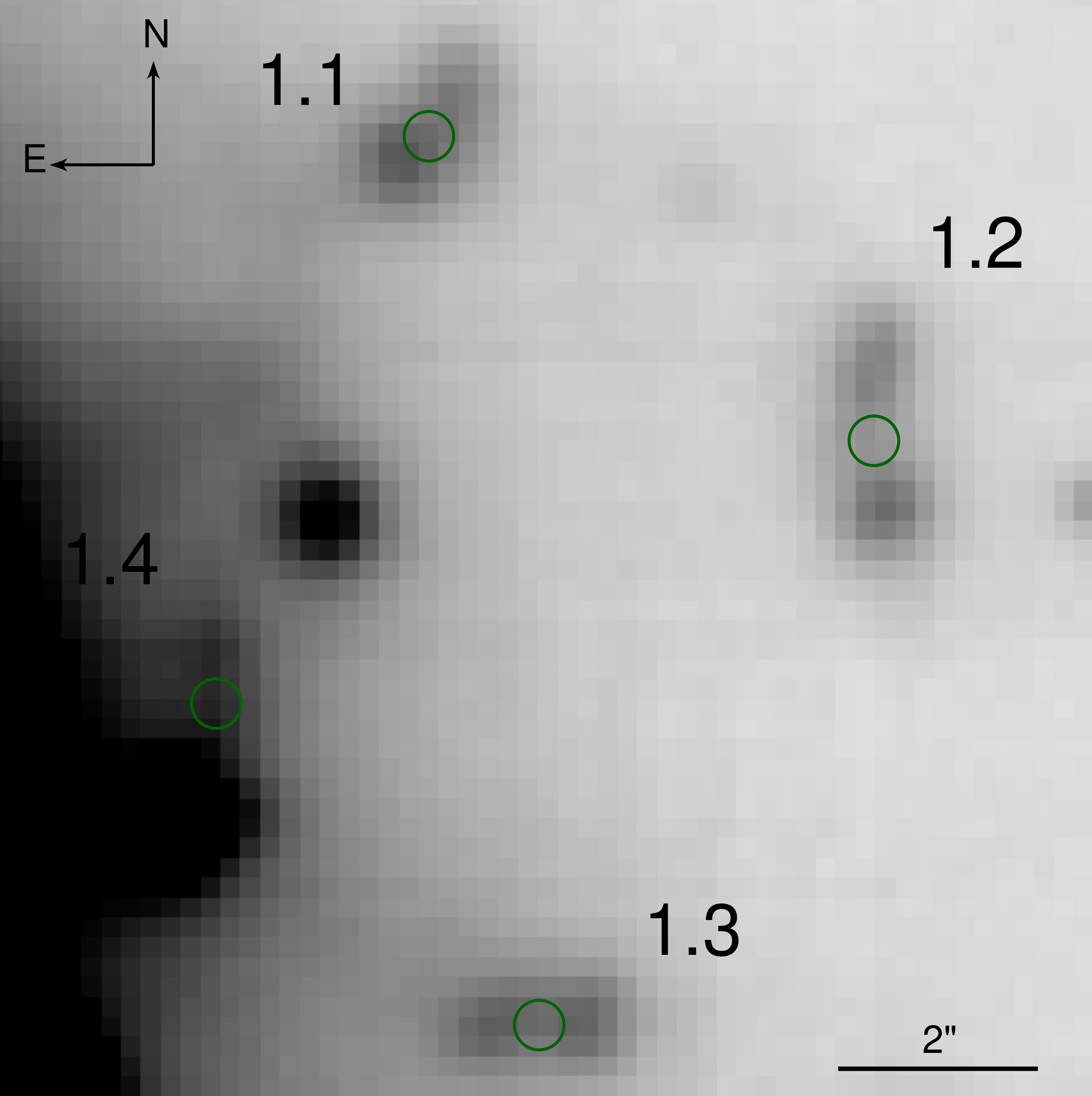}
    \includegraphics[width=0.49\textwidth]{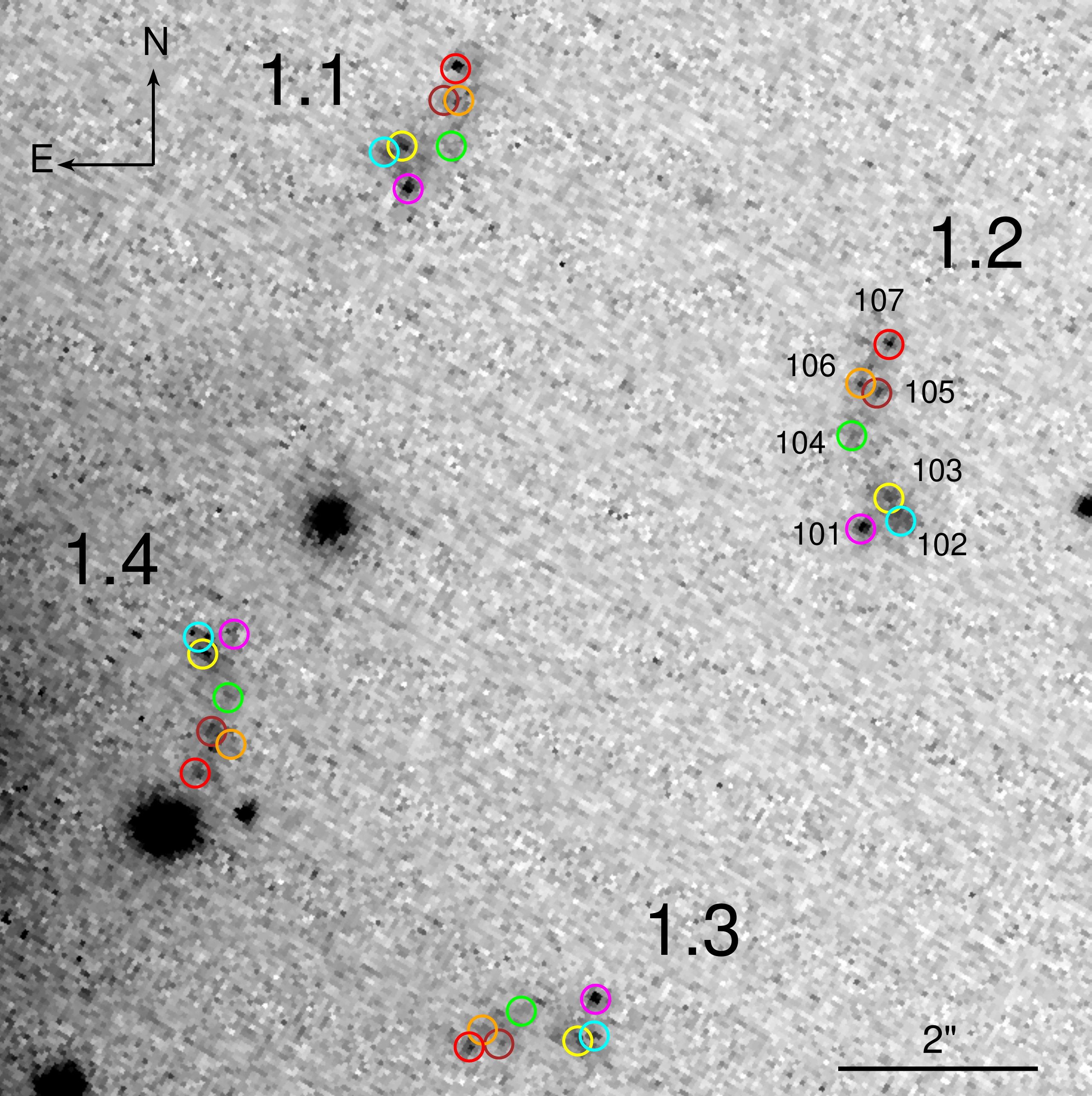}
    \includegraphics[width=0.495\textwidth]{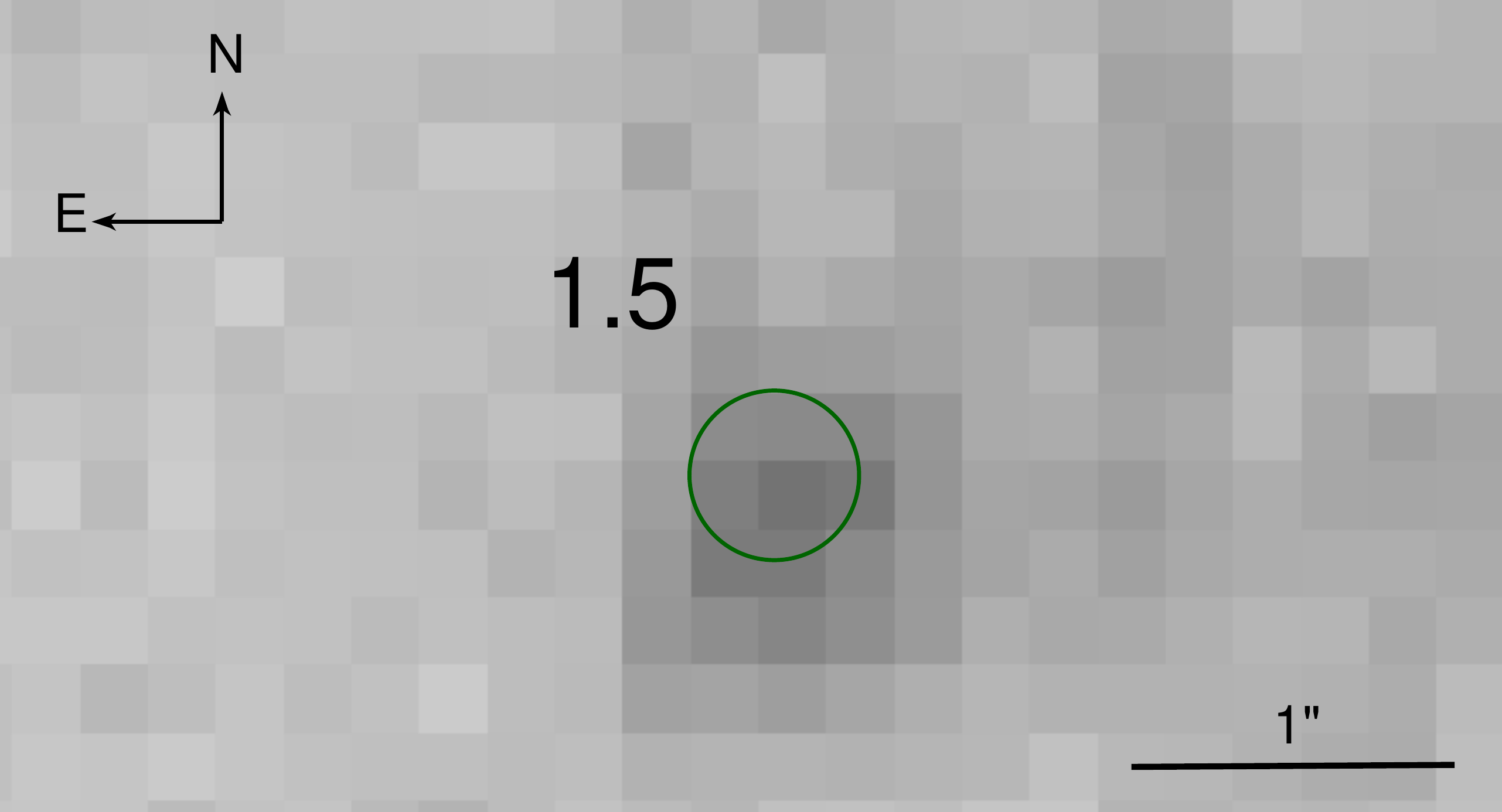}
    \includegraphics[width=0.49\textwidth]{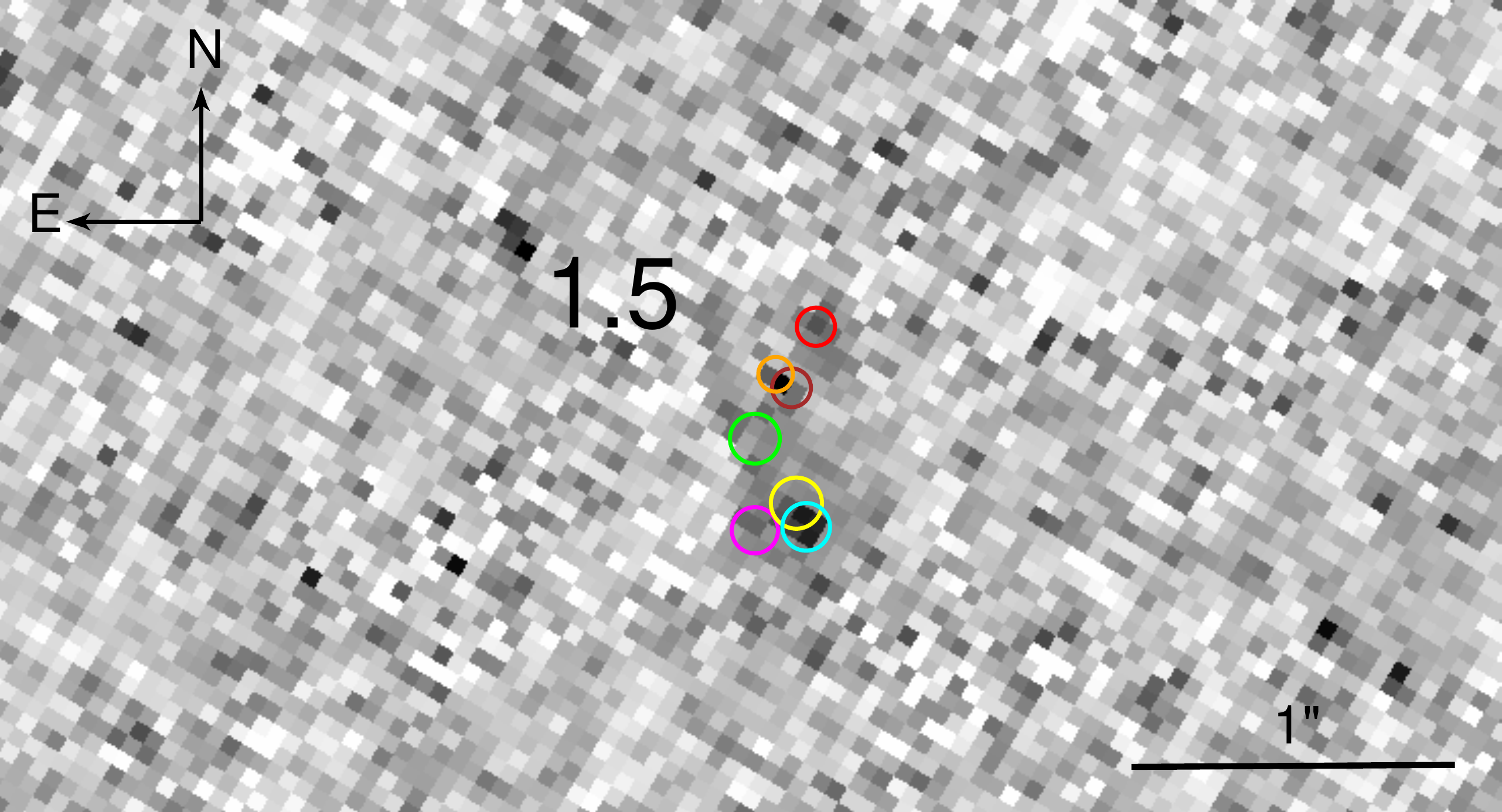}
    \caption{Structure seen in the continuum emission from System 1, at MUSE (left) and HST (right) resolution. While some broad structure is seen in the MUSE data, the \emph{HST} images reveal considerably more complexity: seven distinct clumps (labelled 101 to 107) can be seen in each of the four counterimages. Each of these compact regions can be used as a unique constraint in the lens model.} 
    \label{fig:arc_compare}
\end{figure*}

Thanks to the higher resolution provided by \emph{HST}, we are able to identify individual star-forming knots in two systems (Systems 1 and 2) providing additional constraints to the model (Table~\ref{tbl:sysHiRes}). This substantial advantage of space-based angular resolution is illustrated in Fig.~\ref{fig:arc_compare} which juxtaposes System 1 as seen with MUSE and \emph{HST}. Remarkably, the \emph{HST} imaging of System 2 reveals it to be yet another H-U system (subsequently confirmed by the model itself) bringing the total number of exotic lenses in the cluster to three.  

\begin{table}
  \centering
  \caption{Multiple-image systems. Systems 1, 2, and 10 are H-U configurations.} 
  \label{tbl:Multi-Images}
  \begin{tabular}[t]{lllc}
    \hline
    ID & RA & Dec & $z$\\
    \hline
      1.1 & 69.28763699 & 0.7328107580 & 2.9732\\
      1.2 & 69.28638970 & 0.7319299537 & 2.9732\\
      1.3 & 69.28732442 & 0.7302999442 & 2.9732\\
      1.4 & 69.28835037 & 0.7311433627 & 2.9732\\
      1.5 & 69.29780033 & 0.7290919997 & 2.9732\\
      2.1 & 69.29222312 & 0.7310222552 & 1.9722\\ 
      2.2 & 69.29091637 & 0.7302348672 & 1.9722\\ 
      2.3 & 69.29100445 & 0.7299858009 & 1.9722\\ 
      2.4 & 69.29234516 & 0.7307900313 & 1.9722\\ 
      2.5 & 69.28210829 & 0.7330253619 & 1.9722\\
      3.1 & 69.28927610 & 0.7247680088 & 5.2400\\
      3.2 & 69.29210940 & 0.7248924694 & 5.2400\\
      3.3 & 69.28193799 & 0.7290414295 & 5.2400\\
      4.1 & 69.28970182 & 0.7332842430 & 5.3151\\
      4.2 & 69.29607434 & 0.7282058544 & 5.3151\\
      4.3 & 69.28774638 & 0.7274040572 & 5.3151\\
      4.4 & 69.28335046 & 0.7314971582 & 5.3151\\
      5.1 & 69.29152354 & 0.7308130095 & 3.5296\\
      5.2 & 69.29176084 & 0.7309814052 & 3.5296\\
      6.1 & 69.29032696 & 0.7368201796 & 3.2612\\
      6.2 & 69.28580688 & 0.7363360001 & 3.2612\\
      6.3 & 69.29649552 & 0.7325319083 & 3.2612\\
      7.1 & 69.29658902 & 0.7323622317 & 5.2013\\
      7.2 & 69.29116228 & 0.7365076884 & 5.2013\\
      7.3 & 69.28425585 & 0.7357322308 & 5.2013\\
      8.1 & 69.29351399 & 0.7320022685 & 3.5343\\
      8.2 & 69.29326011 & 0.7325189371 & 3.5343\\
      9.1 & 69.28754874 & 0.7343179871 & 3.5691\\
      9.2 & 69.28568583 & 0.7333683333 & 3.5691\\
     10.1 & 69.28609916 & 0.7308694335 & 6.0196\\
     10.2 & 69.28647389 & 0.7303784567 & 6.0196\\
     10.3 & 69.28814076 & 0.7318643094 & 6.0196\\
     10.4 & 69.28805031 & 0.7325103319 & 6.0196\\
     11.1 & 69.28710600 & 0.7288205549 & 0.9011\\
     11.2 & 69.28683786 & 0.7291805028 & 0.9011\\
     12.1 & 69.2966431  & 0.7301256    & 4.7668\\  
     12.2 & 69.2903813  & 0.7353034    & 4.7668\\
     12.3 & 69.2838421  & 0.7335403    & 4.7668\\
     13.1 & 69.2858521  & 0.7260117    & 6.4425\\
     13.2 & 69.2846311  & 0.7266802    & 6.4425\\
    \hline  
    \end{tabular}
\end{table}

\begin{table}
    \centering
    \caption{Positions of the subclumps of Systems 1 and 2}
    \label{tbl:sysHiRes}
    \begin{tabular}{l|l|l}
    \hline
    ID  &  RA  &  Dec\\
    \hline
    101.1  &  69.287726  &  0.73265278 \\
    101.2  &  69.286460  &  0.73169867 \\
    101.3  &  69.287200  &  0.73038422 \\
    101.4  &  69.288213  &  0.73140660 \\
    101.5  &  69.297811  &  0.72911185 \\
    102.1  &  69.287793  &  0.73275422 \\
    102.2  &  69.286346  &  0.73172222 \\
    102.3  &  69.287203  &  0.73028132 \\
    102.4  &  69.288313  &  0.73139770 \\
    102.5  &  69.297772  &  0.72911443 \\
    103.1  &  69.287741  &  0.73277225 \\
    103.2  &  69.286378  &  0.73178569 \\
    103.3  &  69.287248  &  0.73026690 \\
    103.4  &  69.288299  &  0.73135082 \\
    103.5  &  69.297781  &  0.72913459 \\
    104.1  &  69.287605  &  0.73277164 \\
    104.2  &  69.286482  &  0.73196032 \\
    104.3  &  69.287410  &  0.73035067 \\
    104.4  &  69.288228  &  0.73122689 \\
    104.5  &  69.297814  &  0.72918580 \\
    105.1  &  69.287625  &  0.73289960 \\
    105.2  &  69.286411  &  0.73208179 \\
    105.3  &  69.287469  &  0.73025795 \\
    105.4  &  69.288274  &  0.73113417 \\
    105.5  &  69.297784  &  0.72922743 \\
    106.1  &  69.287583  &  0.73290068 \\
    106.2  &  69.286460  &  0.73210884 \\
    106.3  &  69.287518  &  0.73029797 \\
    106.4  &  69.288219  &  0.73109631 \\
    106.5  &  69.297797  &  0.72923780 \\
    107.1  &  69.287593  &  0.73298939 \\
    107.2  &  69.286379  &  0.73221701 \\
    107.3  &  69.287554  &  0.73025038 \\
    107.4  &  69.288320  &  0.73101626 \\
    107.5  &  69.297763  &  0.72927883 \\
    201.1  &  69.292159  &  0.73114849 \\
    201.2  &  69.290957  &  0.73034018 \\
    201.3  &  69.291104  &  0.72986240 \\
    201.4  &  69.292312  &  0.73071579 \\
    201.5  &  69.282125  &  0.73301792 \\
    202.1  &  69.292121  &  0.73154732 \\
    202.2  &  69.290800  &  0.73049663 \\
    202.3  &  69.291010  &  0.72956682 \\
    202.4  &  69.292545  &  0.73056312 \\
    202.5  &  69.282179  &  0.73299134 \\
    \hline
    \end{tabular}
\end{table}

\section{Lens model}
\label{sec:model}

\subsection{Model setup and results}

When characterizing cluster-scale mass components, we begin with the assumption of a single halo, centred close to the BCG. After finalizing the selection of cluster members and multiple images, and combining this information with the cluster-scale halo, we construct an initial set of model parameters for \textsc{Lenstool} analysis.

To investigate the complex and highly structured morphology of H-U Systems 1 and 2, we test two different setups:\ a ``low-res'' version based on the MUSE-identified centroid of each lensed image (as shown in Fig. \ref{fig:data}) and a ``high-res'' version that treats individual stellar knots as separate constraints (that is, we replace the System 1 and System 2 constraints listed in Table \ref{tbl:Multi-Images} with those presented in Table \ref{tbl:sysHiRes}). In both setups we begin model optimization using the single-cluster-halo assumption mentioned previously. However, after several sampling iterations we find that the resulting model does not adequately fit the lensing constraints, with particularly large residuals ($> 1\arcsec$) for images 3.1 and 3.2, as well as image 4.3. To improve the fit, we therefore add a second DM component in the neighbourhood of these constraints, but give \textsc{Lenstool} freedom to adjust its position by placing a large, uniform prior on this component's initial location ($\pm 10\arcsec$ in each coordinate). Including this extra component results in a significantly improved fit in both the low- and high-res versions of the model. Its final position lies almost due south from the primary cluster halo, separated by $\sim$15\arcsec\ (65 kpc at the cluster redshift). Because this position does not correspond to any bright object or obvious galaxy overdensity, we subsequently name the component the ``dark clump''. Overall, the dark clump is moderately massive, with a central velocity dispersion comparable to that of the BCG. However, it is more elliptical (significantly so in the best-fit hi-res model) and flatter, with a core radius of $\sim$15 kpc. The physical nature of this additional halo remains to be determined:\ it could, for example, represent an asymmetry in the mass distribution of the central halo \citep[e.g.][]{taylor2017,massey2018}, or given its location near the BCG, it may be a diffuse matter overdensity associated with excess intra-cluster light \citep[e.g.][]{Mahler2023} that is not visible in current shallow imaging data. While outside the scope of this work, it nonetheless poses an interesting question for future analysis.

We list the associated best-fit parameters of each setup in Table \ref{tbl:ModParams}. The final model fits of the lensing constraints are excellent, with rms residuals of $\sim 0.3\arcsec$, underlining the benefits of having a high density of constraints, which in this case is driven by the increased number of images provided by the H-U systems. The final parameters of both model setups are in good agreement (typically consistent within $1\sigma$ uncertainty), although we note that the high-res version has a slightly lower rms. Therefore, when discussing model properties and derived features, reference is to the high-res model unless otherwise stated. 

\begin{table*}
  \centering
  \caption{Lens Models and Best-Fitting Parameters}
  \label{tbl:ModParams}
  \begin{tabular}{lcrrrrrrr}
    \hline
    Model Name & Component & $\Delta\alpha^{\rm ~a}$& $\Delta\delta^{\rm ~a}$ & $\varepsilon^{\rm ~b}$ / $\gamma^{\rm ~c}$ & $\theta$  / ~~ $\theta_{\gamma} ^{\rm ~c}$ & $r_{\rm core}$ & $r_{\rm cut}$ & $\sigma_0$\\
    (Fit Statistics)$^{\rm ~d}$ &  & (\arcsec) & (\arcsec) & () ~~/ () & ($\deg$) / ($\deg$) & (kpc) & (kpc) & (km s$^{-1}$)\\
    \hline
    Low-res & Cluster Halo & $  0.17^{+  0.05}_{ -0.04}$ & $  0.71^{+  0.09}_{ -0.12}$ & $ 0.47^{+ 0.01}_{-0.02}$ &   $-72.7^{+  0.5}_{ -0.4}$ & $ 49.1^{+  0.4}_{ -1.7}$ & $[800.0]^{\rm ~e}$ & $930^{+3}_{-15}$ \\[3pt]
    rms = 0\farcs35 & BCG & $[ 0.00]$ & $[  0.00]$ & $[0.37]$ & $[-70.0]$ & $[  0.15]$ & $ 47.6^{+  4.5}_{ -3.4}$ & $269^{+6}_{-4}$ \\[3pt]
    $\chi^2/\nu$ = 2.22 & Dark Clump & $ -0.33^{+  0.35}_{ -2.74}$ & $-16.56^{+  2.09}_{ -1.26}$ & $ 0.88^{+ 0.01}_{-0.08}$ & $138.1^{+ 9.5}_{-10.3}$ &  $43.2^{+0.5}_{-5.9}$ & $[800.0]$ & $193^{+44}_{-2}$ \\[3pt]
    $\log~(\mathcal{L}) = -2.74$ & L$^{*}$ galaxy &  & & & & $[0.10]$ & $ 14.7^{+ 1.0}_{ -0.1}$ & $162^{+4}_{-20}$\\[3pt]
    $\log~(\mathcal{E}) = -80.62$ & Ext.\ Shear & & & $  0.04^{+  0.01}_{ -0.01}$& $  1.1^{+ 13.1}_{ -2.1}$& & & \\
    \hline
    High-res & Cluster Halo & $  0.15^{+  0.08}_{ -0.05}$ & $  1.54^{+  0.25}_{ -0.18}$ & $ 0.44^{+ 0.02}_{-0.02}$ &   $-73.3^{+  0.5}_{ -0.6}$ & $ 43.3^{+  1.7}_{ -2.0}$ & $[800.0]$ & $888^{+12}_{-11}$ \\[3pt]
    rms = 0\farcs29 & BCG & $[ 0.00]$ & $[  0.00]$ & $[0.37]$ & $[-70.0]$ & $[  0.15]$ & $ 36.9^{+  3.9}_{ -11.4}$ & $262^{+7}_{-9}$ \\[3pt]
    $\chi^2/\nu$ = 1.10 & Dark Clump & $ -0.10^{+  0.03}_{ -0.26}$ & $-13.35^{+  0.50}_{ -0.59}$ & $ 0.83^{+ 0.09}_{-0.08}$ & $-33.6^{+ 4.3}_{-4.8}$ &  $15.7^{+3.8}_{-4.9}$ & $[800.0]$ & $214^{+16}_{-18}$ \\[3pt]
    $\log~(\mathcal{L}) = -19.14$ & L$^{*}$ galaxy &  & & & & $[0.10]$ & $ 29.4^{+ 4.1}_{ -4.0}$ & $150^{+6}_{-1}$\\[3pt]
    $\log~(\mathcal{E}) = -36.93$ & Ext.\ Shear & & & $  0.06^{+  0.01}_{ -0.01}$& $  3.1^{+  1.6}_{ -2.6}$& & & \\
\hline
\end{tabular}
\medskip\\
$^{\rm a}$ $\Delta\alpha$ and $\Delta\delta$ are measured relative to the reference coordinate point: ($\alpha$ = 69.289688, $\delta$ = 0.731141)~~~~~~~~~~~~~~~~~~~~~~~~~~~~~~~~~~~~~~~~~~~~~~~~~~~~~~~~~~\\[1pt]
  $^{\rm b}$ Ellipticity ($\varepsilon$) is defined to be $(a^2-b^2) / (a^2+b^2)$, where $a$ and $b$ are the semi-major and semi-minor axes of the ellipse~~~~~~~~~~~~~~~~~~~~~~~\\[1pt]
   $^{\rm c}$ In the shape parameters columns (5 and 6) components representing external shear are described by $\gamma$ and $\theta_{\gamma}$, while other~~~~~~~~~~~~~~~~~~~~~~~ \\components are described by the usual ellipticity terms ($\varepsilon$ and $\theta$).~~~~~~~~~~~~~~~~~~~~~~~~~~~~~~~~~~~~~~~~~~~~~~~~~~~~~~~~~~~~~~~~~~~~~~~~~~~~~~~~~~~~~~~~~~~~~~~~~~~~~~~~~~~~~~\\[1pt]
   $^{\rm d}$ Statistics notes: $\mathcal{L}$ represents the model likelihood and $\mathcal{E}$ the model evidence.~~~~~~~~~~~~~~~~~~~~~~~~~~~~~~~~~~~~~~~~~~~~~~~~~~~~~~~~~~~~~~~~~~~~~~~~~~~~~~~~~~~~~~~~~~~~~~~\\[1pt]
   $^{\rm e}$ Quantities in brackets are fixed parameters~~~~~~~~~~~~~~~~~~~~~~~~~~~~~~~~~~~~~~~~~~~~~~~~~~~~~~~~~~~~~~~~~~~~~~~~~~~~~~~~~~~~~~~~~~~~~~~~~~~~~~~~~~~~~~~~~~~~~~~~~~~~~~~~~~~~~~~~~~~~~~~~~~~~~~
\end{table*}

\subsection{Analysis and mass distribution}

\begin{figure}
    \centering
    \includegraphics[width=0.49\textwidth]{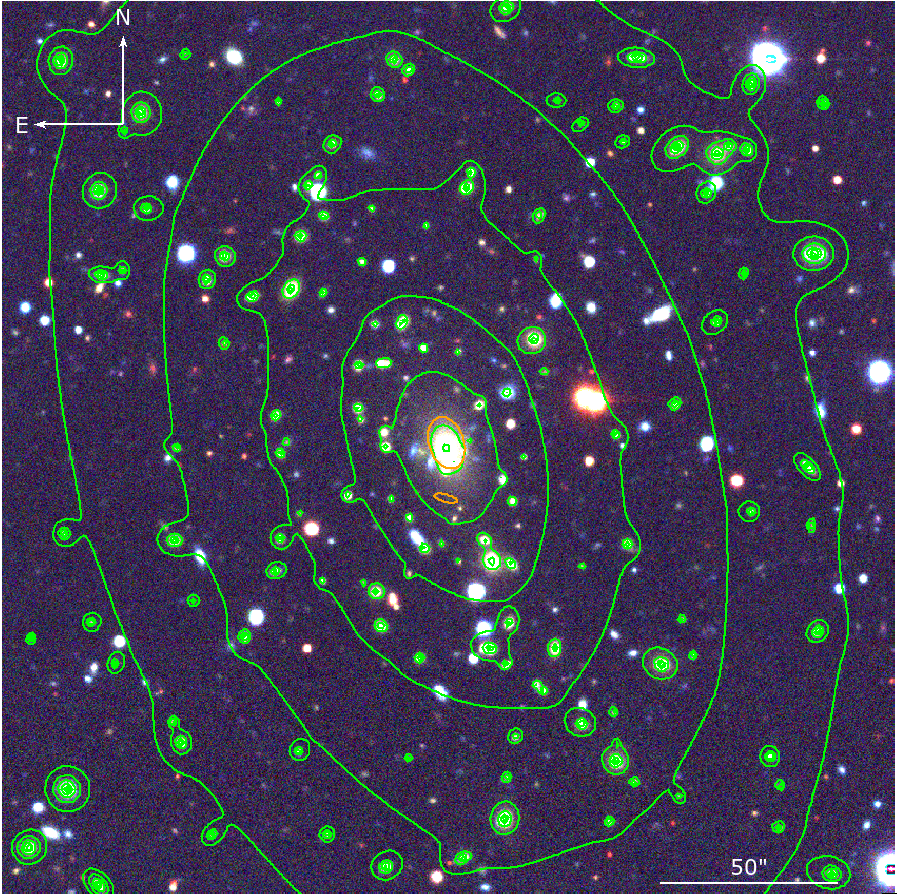}
    \caption{Mass contours of the best-fit ``high-res'' lens model, overlaid on DECam imaging. The overall distribution is highly elliptical out to large radii and has an orientation aligned with the BCG. Ellipses showing the centroid and PA of the two large-scale mass components (the cluster halo and ``dark clump'') are shown in orange. We note that the ellipses do not show the full extent of each mass component:\ instead, they are scaled by the best-fit velocity dispersion ($\sigma_0$) and ellipticity ($\varepsilon$) parameters of each component, to show their shapes and relative contributions to the mass model.}
    \label{fig:massMap}
\end{figure}

\begin{figure*}
    \centering
    \includegraphics[width=0.32\textwidth]{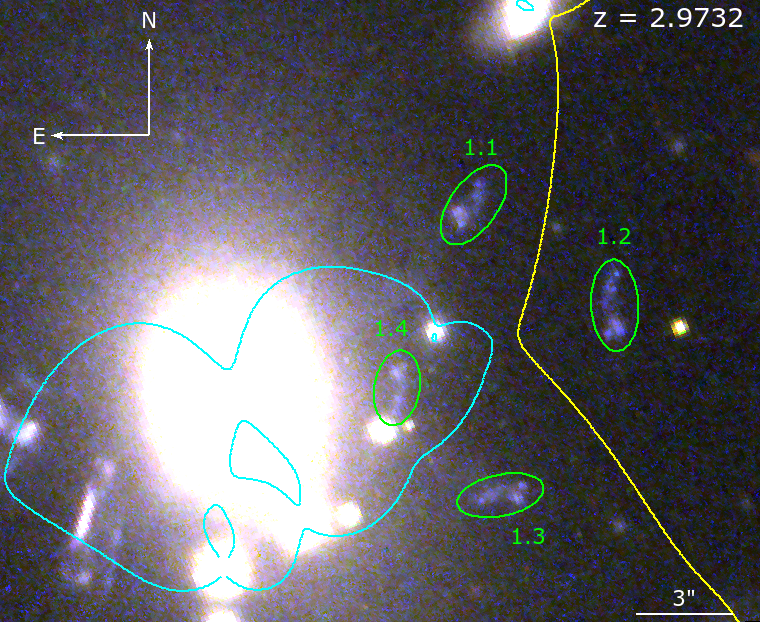}
    \includegraphics[width=0.32\textwidth]{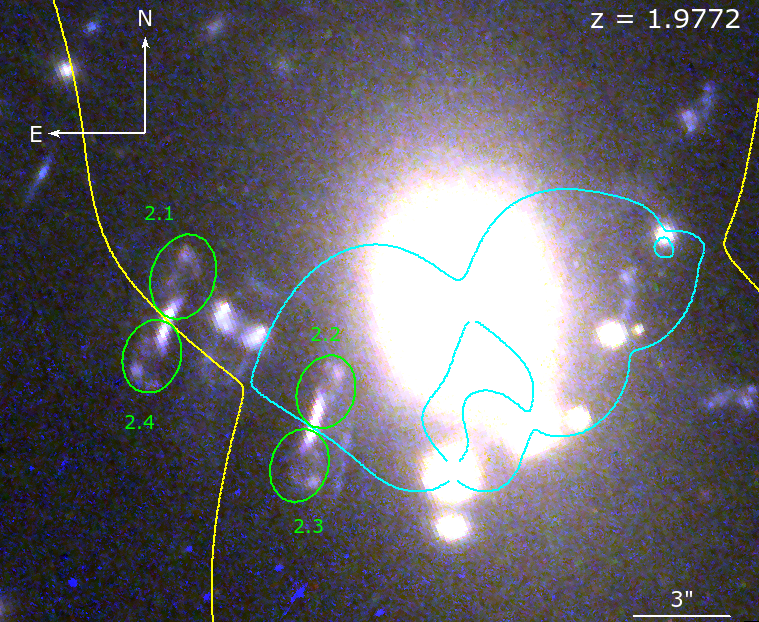}
    \includegraphics[width=0.32\textwidth]{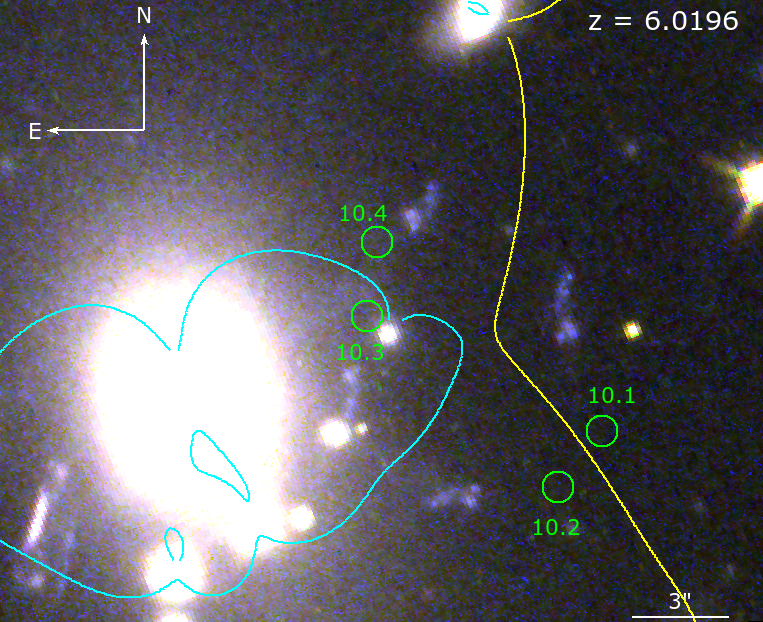}
    \caption{Cutouts of the H-U lens systems overlaid on multi-band \emph{HST} imaging. We present from left to right:\ System 1, System 2, and System 10. In each panel, the identified multiple images are labelled by green ellipses, while the tangential (radial) critical curve at the source redshift is shown as a yellow (cyan) line. Unlike traditional multiple-image systems, H-U galaxies are located close to both critical curves. As a result, they experience a more uniform magnification over their entire surface.}
    \label{fig:crit_curves}
\end{figure*}

The two largest mass components in the model are the cluster halo and the BCG, although a non-negligible external shear term ($\gamma \sim 0.05$) suggests the presence of additional mass contributions in the vicinity of the main field. We find both the cluster-scale and the BCG components to be moderately elliptical ($\varepsilon \sim 0.4$), with nearly spatially coincident centroids (to within $2\arcsec$) and closely aligned position angles ($\theta \sim -70\deg$).  This implies that the overall mass distribution will also have moderate to high ellipticity, as is confirmed by the 2D surface mass map  (Fig.~\ref{fig:massMap}) which shows clearly elliptical mass contours originating from the cluster centre that maintain their shape even out to high radii. The large elongation helps explain the large number of H-U systems in this cluster since, in elliptical mass distributions, the two primary critical curves (one radial and one tangential) will be forced close together in areas lying along the minor axis. At high enough ellipticity, these critical lines will nearly touch, a key indicator for exotic lenses. Tracing the critical curves at the redshifts of each H-U candidate, we do indeed see the characteristic close pass of the two lines near each multiple image (Fig. \ref{fig:crit_curves}), providing further proof of their nature. 

\begin{figure*}
    \centering
    \includegraphics[width=0.49\textwidth]{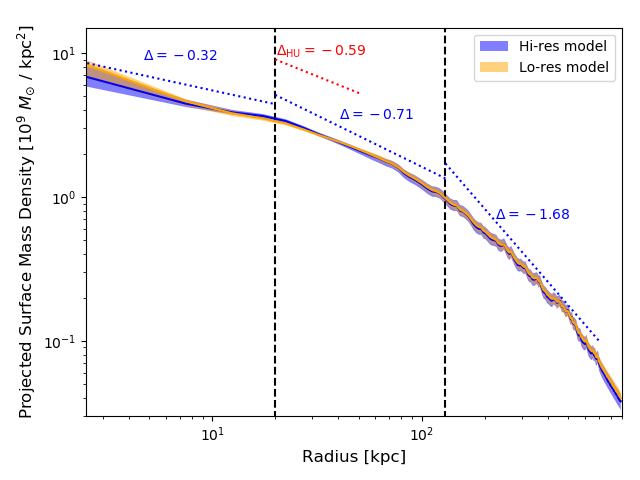}
    \includegraphics[width=0.49\textwidth]{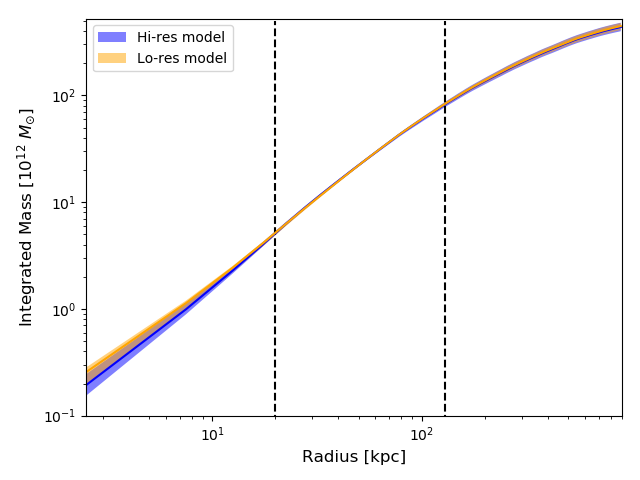}
    \caption{Left:\ Radially averaged surface mass density profile of RXJ0437, using the hi- and lo-res models described in Section \ref{sec:model}. Profiles are measured with elliptical apertures, where the ellipticity ($\varepsilon$ = 0.42) and PA ($\theta = 18.4$ deg) are determined from the average shape of the mass contours seen in Fig.~\ref{fig:massMap}. Solid coloured lines show the best-fit profile, while fainter shaded areas provide 3$\sigma$ error estimates. The area encompassing the strong-lensing constraints is bracketed by the vertical dashed lines, making it the most well-constrained segment of the plot. To quantify the profile shape, we measure its logarithmic slope over three regions (blue dotted lines), finding a steep increase as a function of radius. To highlight the continually changing nature of the profile, we also measure the slope over the region containing only H-U lens systems (red dotted line), measuring a shallower value compared to the full multiple-image region. Right:\ Integrated density profile, showing the enclosed total cluster mass as a function of radius.} 
    \label{fig:profiles}
\end{figure*}

When radially averaging the map (Fig. \ref{fig:profiles}, left), we observe a mass density profile that is relatively flat in the central core and then gradually steepens at larger radii. To quantify the shape of the distribution we measure its logarithmic slope, splitting the total profile into three separate regions based on distance. We define the slope as $\Delta \equiv \log({\Sigma(r)}$) / $\log(r)$, where $\Sigma(r)$ is the measured (2D) surface mass density at a given radius, as determined by lensing. In the innermost region ($r < 20$ kpc; the radius set by the multiple image closest to the centre) we measure the flattest slope, $\Delta = -0.32$. At intermediate radii ($20$ kpc $< r < 130$ kpc; the region containing all observed multiple-image systems, and thus the best-constrained segment of the profile) the slope steepens to $\Delta = -0.71$. However, if we instead limit our measurement to only the H-U region ($20 < r < 52$ kpc) we find a slope $\Delta = -0.59$, highlighting the variability of the profile in this regime. Finally, at the largest radii ($r > 130$ kpc; the area beyond the furthest multiple-image constraint) we measure an even steeper slope of $\Delta = -1.68$, although this value may be biased by edge effects at the mass-map boundary ($\pm700$ kpc from the centre of the BCG). Strictly speaking, only the slope in the intermediate region is based on data, as the inner and outer sections of the profile do not contain any lensing constraints. However, due to the smoothly varying nature of our parametric model, we expect these projected slopes to behave similarly to the actual profiles, especially at radii close to the zone boundaries. We note, too, that the area of the unconstrained inner profile is considerably smaller in RXJ0437 than in most clusters, thanks to the presence of the H-U images. Integrating the mass profile, we also measure the total mass as a function of radius (Fig. \ref{fig:profiles}, right), and find values of ($5.09 \pm 0.14$) $\times 10^{12} M_{\odot}$ in the inner 20 kpc, ($8.22 \pm 0.44$) $\times 10^{13} M_{\odot}$ through the edge of the multiple-image region, and ($4.37 \pm 0.43$) $\times 10^{14} M_{\odot}$ over the full extent of the mass map. 

\section{H-U Source Galaxy Properties}
\label{sec:source}

Due to their closeness to the critical curves, images of source galaxies are significantly magnified in a H-U lensing configuration, providing a valuable window into the physical properties of the lensed background sources. In this section we investigate aspects of the H-U galaxies in RXJ0437, focusing primarily on System 1, but also commenting on Systems 2 and 10.  

\subsection{Spectral features}

\begin{figure}
    \centering
    \includegraphics[width=0.45\textwidth]{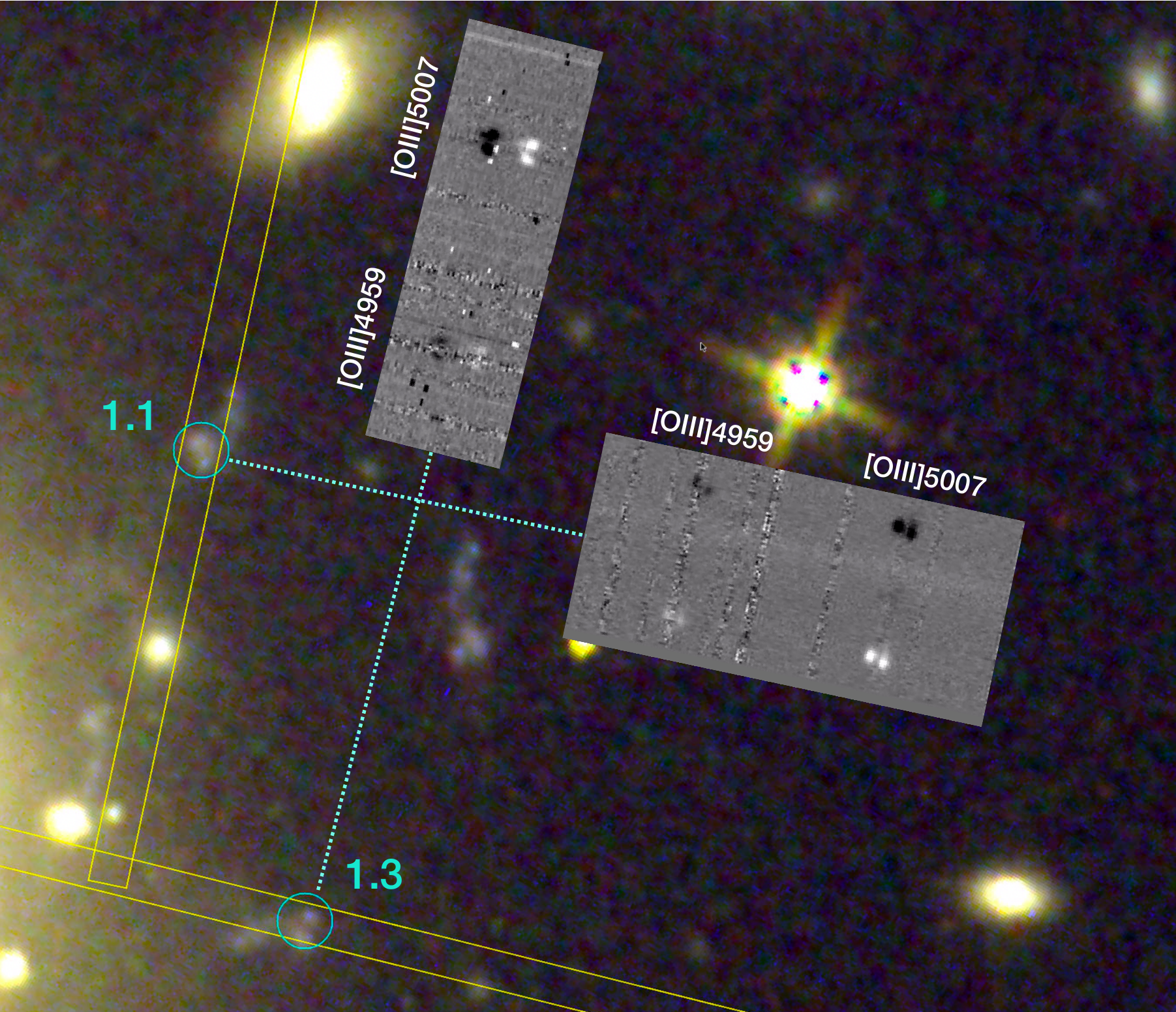}
    \caption{Slit positions and orientations as well as resulting 2D spectrum of the [OIII] emission from images 1.1 and 1.3 as observed with Keck/MOSFIRE in the K band; different dither offsets of 5 and 1.5\arcsec\ were used for the two components, resulting in a wider separation of the positive and negative signal for image 1.1. All [OIII] lines are detected as doublets. The spatial offset between the two components apparent in the 2D spectrum of image 1.1 suggests that the two velocity components are uniquely associated with the two knots 101 and 102/103 in System 1 (see Fig.~\ref{fig:arc_compare} and Table~\ref{tbl:sysHiRes}).}
    \label{fig:mosfire2d}
\end{figure}

\begin{figure*}
    \centering
    \includegraphics[width=\textwidth]{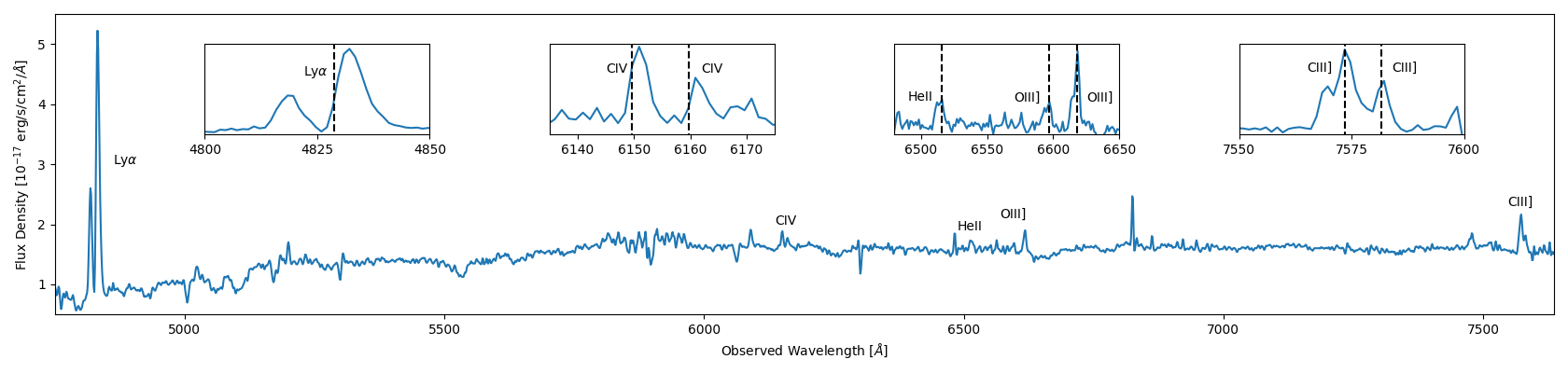}
    \includegraphics[width=0.49\textwidth]{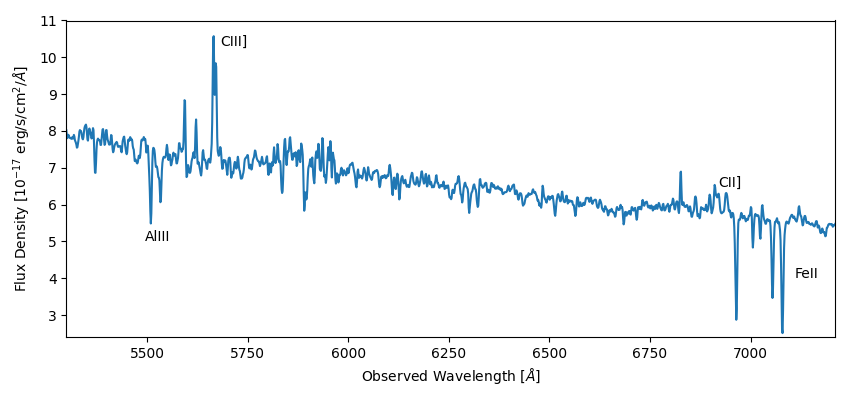}
    \includegraphics[width=0.49\textwidth]{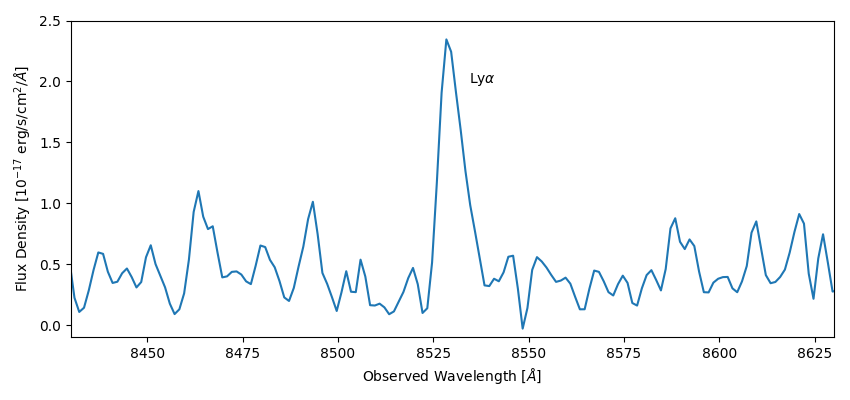}
    
    \caption{Extracted MUSE spectra of the H-U galaxies in the RXJ0437 field. Top: System 1 at $z = 2.9732$. The full panel shows the extraction over the Ly$\alpha$ ring, while the inset panels are extracted from the continuum regions to better highlight the more localized line features. Bottom left: System 2 at $z  = 1.9722$, with a steep UV continuum and strong CIII] emission line. Bottom right: System 10 at $z = 6.0196$; only a moderate Ly$\alpha$ emission feature can be seen, and the galaxy has no clear continuum in \emph{HST} imaging.}
    \label{fig:lyaSpec}
\end{figure*}

The extracted optical spectrum of System 1 (Fig.~\ref{fig:lyaSpec}) shows a bright UV continuum and several prominent line features. The strongest of these is Lyman-$\alpha$ (Ly$\alpha$), although we also see clear evidence of other high-ionization emission features, such as CIII] (1907,1909 \AA), CIV (1548,1550 \AA), OIII] (1661,1666 \AA), and HeII (1640 \AA). In addition, the NIR spectrum (Fig.~\ref{fig:mosfire2d}) shows pronounced [OIII] (4959,5007 \AA) emission. Taken together, the presence of these lines suggests that the galaxy is young and experiencing a period of enhanced star formation \citep[e.g.,][]{patricio2016,erb2018}. 

Intriguingly, the Ly$\alpha$ feature has a distinct double-peaked appearance, with a bright (red-side) component observed at $\lambda = 4832$ \AA\ and a fainter "blue-bump" feature offset by 14\AA, at $\lambda = 4818$ \AA. Because of the complex spectral profile of the Ly$\alpha$ emission, we do not use either component of the line to measure the galaxy redshift; instead we opt for CIII], which is unaffected by potential inflow/outflow movements of the Ly$\alpha$ gas. Centring on the peaks of the CIII] doublet, we measure a redshift of $z = 2.9732$, which we use as the systemic value in the lens model. Compared to this measurement, the red-side Ly$\alpha$ peak is offset by 260 km s$^{-1}$, while the blue component is  shifted by $-640$ km s$^{-1}$. Clear velocity substructure in the brightest part of System 1 is also observed in the [OIII] lines detected in the NIR with Keck/MOSFIRE and shown in Fig.~\ref{fig:mosfire2d}; the spectral separation of the line components of about 13\AA\ is the same as seen in the Ly$\alpha$ line and can, thanks to the orientation of the MOSFIRE slit, be attributed to the peculiar motion of subsystem 101 relative to the 102/103 complex.

\begin{figure*}
    \centering
    \includegraphics[width=0.33\linewidth]{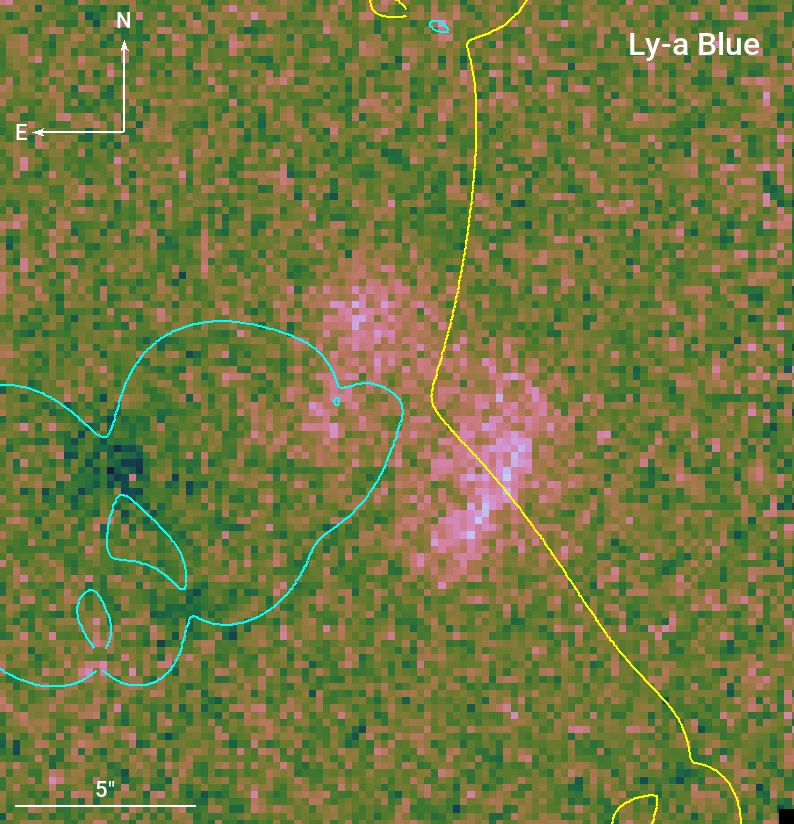}
    \includegraphics[width=0.33\linewidth]{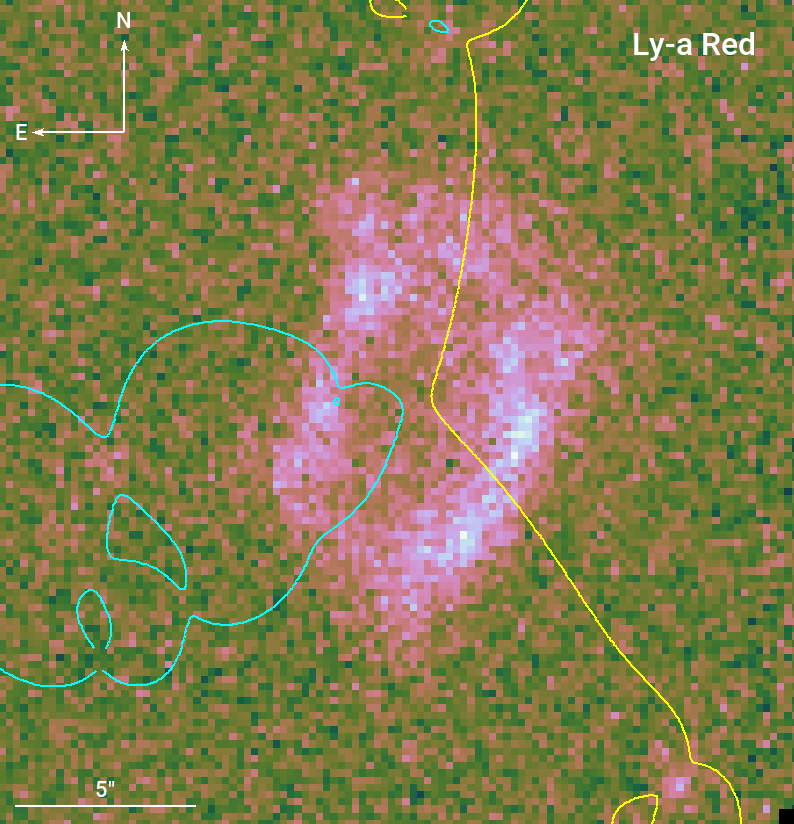}
    \includegraphics[width=0.33\linewidth]{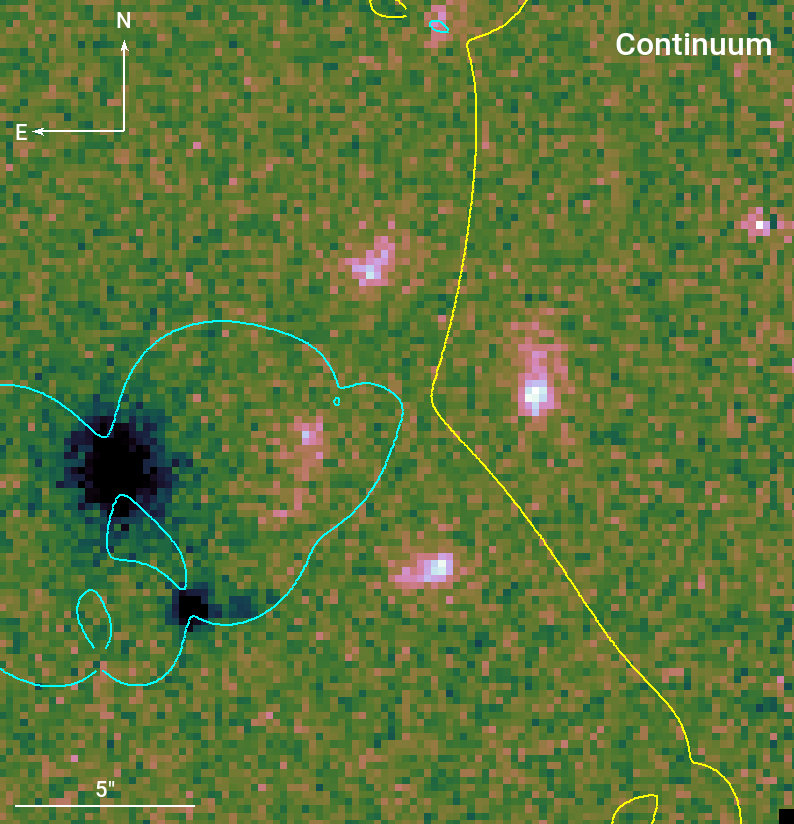}
    \caption{Size comparison between different emission regions in System 1, as seen in MUSE data. The left two panels highlight the complex nature of the double-peaked Ly$\alpha$ gas, while the right panel displays the stellar continuum. Compared to the continuum, the Ly$\alpha$ region is significantly larger, though there is a noticeable difference between the sizes of the main red-side emission and the smaller ``blue bump'' (Fig.~\ref{fig:lyaSpec}). The flux peaks of each emission region are also physically offset by as much as $\sim1\arcsec$. }
\label{fig:lyaExtent}
\end{figure*}

As the minimum between the two Ly$\alpha$ peaks is itself blueshifted from the systemic redshift by $\sim$180 km s$^{-1}$, the two peaks have an average velocity of $\pm$450 km s$^{-1}$ relative to the galaxy when this additional bulk motion is accounted for. Spatially, we find a red/blue asymmetry in the position of the Ly$\alpha$ emission (Fig.~\ref{fig:lyaExtent}), with the red side extending further than the blue (creating the ring-like structure seen in the MUSE data) and the peak of its emission located $\sim 1\arcsec$ further from the critical curves. Compared to the continuum emission, however, both the red- and blue-side Ly$\alpha$ emissions extend much closer to the critical lines, therefore experiencing considerably higher lensing magnification.

While not as complex as System 1, the source galaxies of Systems 2 and 10 still have identifiable spectra, with System 2 ($z = 1.9722$) showing strong CIII] and [OIII] emission, as well as several metal absorption lines embedded within a steeply declining UV continuum, and System 10 ($z = 6.0196$) featuring a single, moderately bright Ly$\alpha$ line. Like System 1, the emission features of these galaxies extend over both the radial and tangential critical curves, giving rise to the observed H-U configurations. However, in the case of System 2, the galaxy continuum also touches the critical curves, hyper-resolving stellar features in these areas.

\subsection{Source plane reconstructions}
\begin{figure*}
  \centering{
  \includegraphics[width=0.485\linewidth]{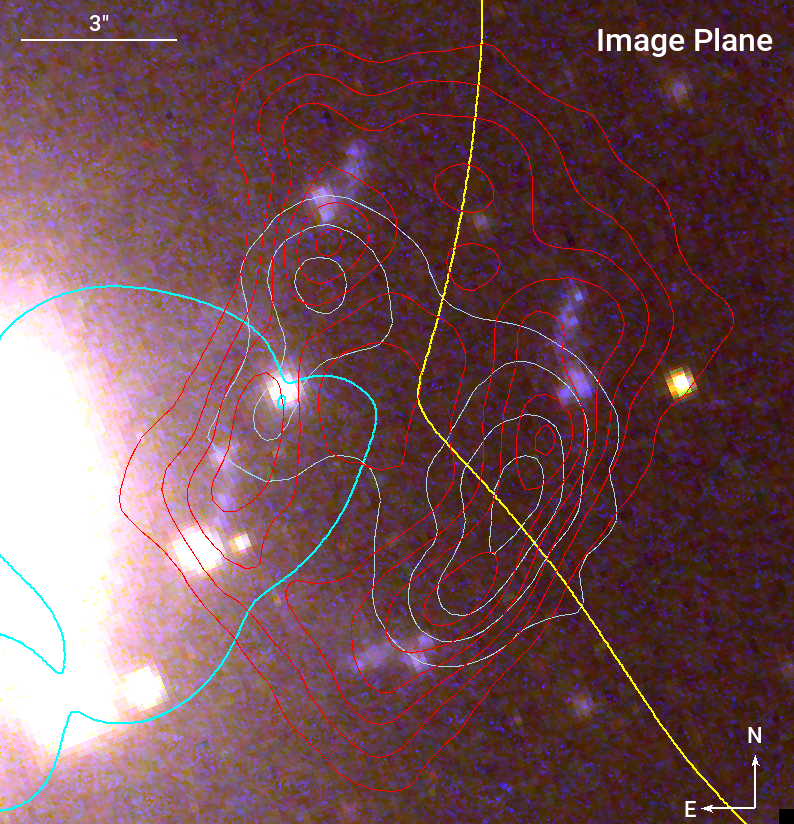}
  \includegraphics[width=0.485\linewidth]{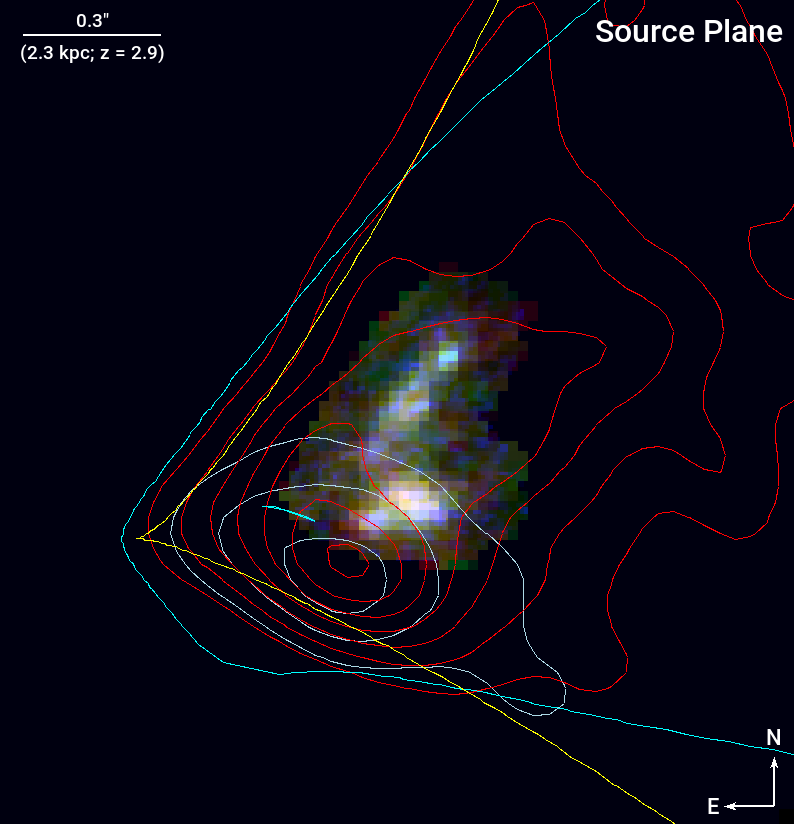}
  }
  \caption{Left: Zoom in of the System\,1 H-U region. The RGB image shows the positions of the UV continuum, while the red- and blue-side components of the Ly$\alpha$ emission (Fig.\,\ref{fig:lyaExtent}) are overlaid as colored contours. Right: Source-plane reconstruction of the galaxy.  The model caustic curves are shown as yellow and cyan lines, corresponding to the critical curves. While the reconstructed continuum region (coloured pixels) is just contained within the caustics, the Ly$\alpha$ halo (coloured contours) extends beyond it, giving rise to a complete ring pattern.}
  \label{fig:ringSource}
  \end{figure*}

Using the best-fit model to remove the effects of lensing, we reconstruct the undistorted (source plane) appearance of the H-U galaxies with \textsc{lenstool}, giving a clearer picture of their intrinsic sizes and shapes. Doing so for System 1 (Fig.~\ref{fig:ringSource}), we see that the source-plane continuum (right panel) looks remarkably similar to its lens-plane counterpart (left panel). Similar results can be seen in the Ly$\alpha$ emission, but because the Ly$\alpha$ flux ultimately merges along the critical curves (meaning that only a fraction of the full component is seen in the image) the differences between planes is more noticeable. The fact that the galaxy shape is only mildly altered by lensing is a particular feature of H-U systems:\ because the source lies close to both the radial and tangential caustic lines (the source-plane equivalent of the critical curves), lensing distortions affect both axes of the galaxy nearly equally. Thus, rather than a shearing over one preferred axis, the galaxy simply undergoes a (roughly) circular, uniform magnification.  As a result, the observed lens-plane appearance of the galaxy provides distinct information about its resolved structure along the full, 2D extent of the galaxy.  

The major axis of the System 1 continuum region spans $\sim$0.4\arcsec\ in the source plane, corresponding to a physical size of $\sim$3.1 kpc at the systemic redshift. The individual luminous substructures identified in Fig.~\ref{fig:arc_compare} are also clearly visible, having diameters of 120 - 240 pc each, when accounting for the instrumental PSF. Qualitatively, we see a slight colour gradient over the galaxy, with clumps in the north (above the small gap in the centre) appearing bluer in F814W/F110W/F140W colour space than those in the south. The intensity of the emission lines relative to the continuum is also up to 50\% stronger in the southern clumps, though we do not notice any significant difference in line widths.

\begin{figure*}
    \centering
    \includegraphics[width=0.485\linewidth]{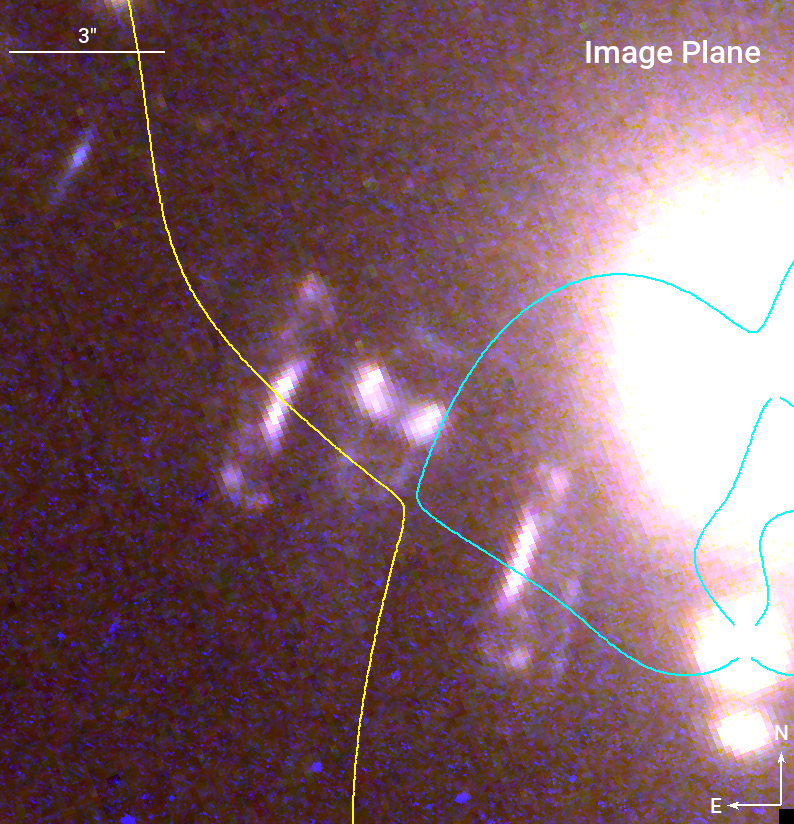}
    \includegraphics[width=0.485\linewidth]{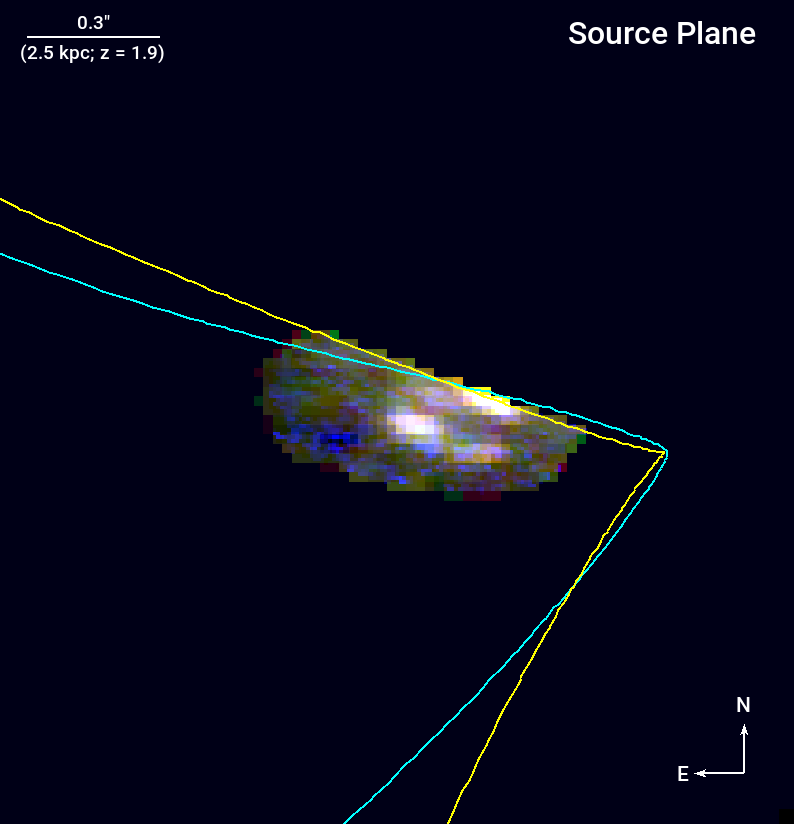}
    \includegraphics[width=0.485\linewidth]{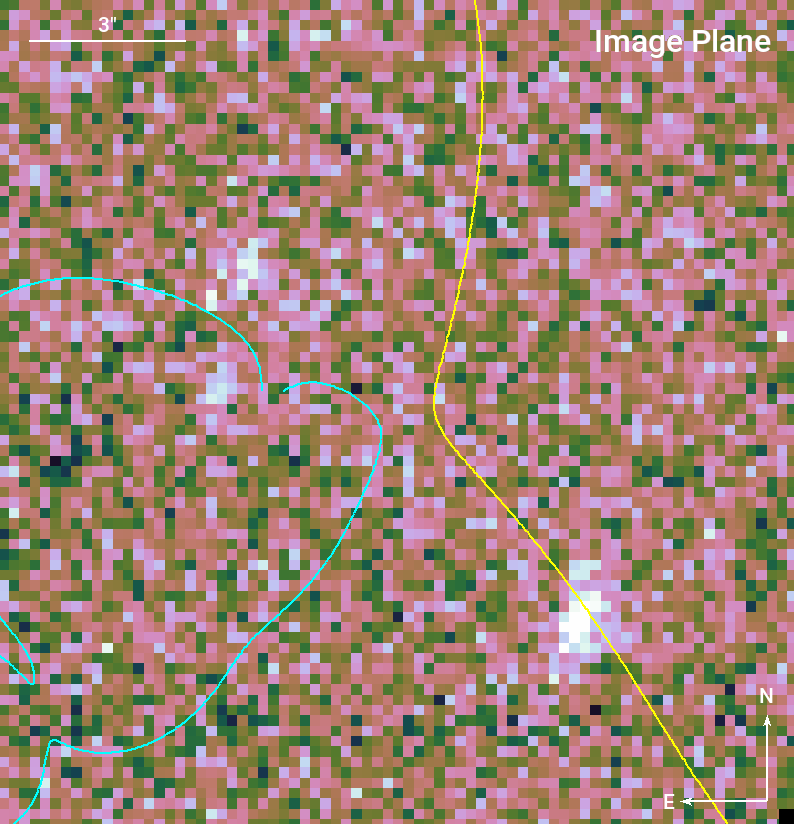}
    \includegraphics[width=0.485\linewidth]{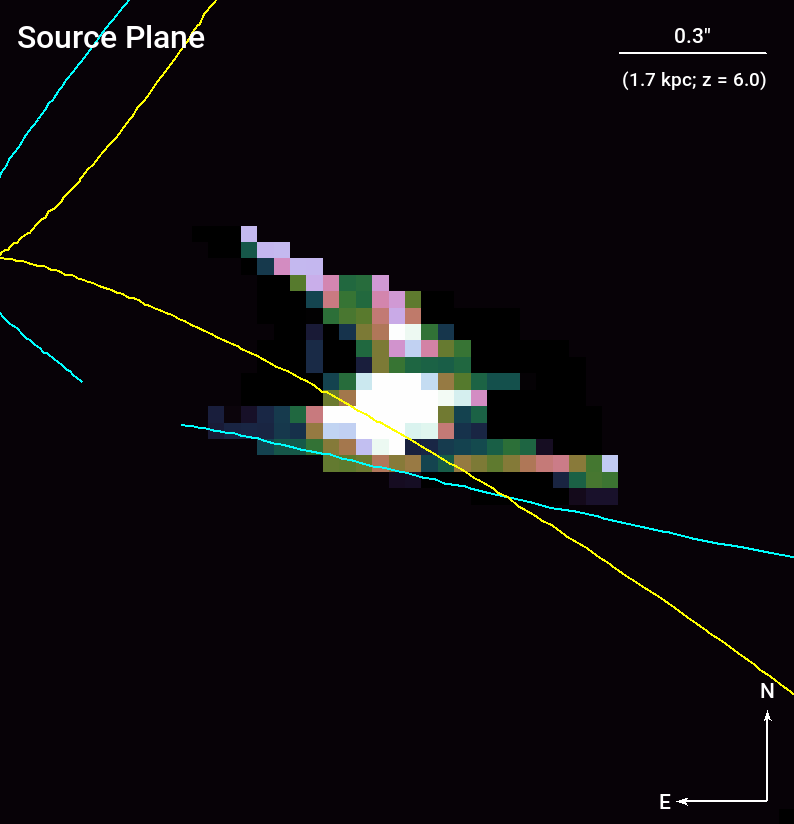}
    \caption{Same as Fig.~\ref{fig:ringSource}, but for Systems 2 and 10, respectively.}
    \label{fig:otherSources}
\end{figure*}

We present additional reconstructions of the remaining H-U galaxies (Systems 2 and 10) in Fig.~\ref{fig:otherSources}, gaining insight into their structures as well. Starting with System 2, we find that the continuum of the galaxy merges along the critical lines (meaning they extend beyond the source-plane caustic lines), resulting in an incomplete reconstruction and a magnification enhancement that is not as perfectly uniform. This is similar to the Ly$\alpha$ component of System 1. Nonetheless, we still identify distinct features along both axes of the galaxy (spanning a distance of 2.4 kpc along its non merging axis), including additional luminous substructure clumps at the $\sim$ 300 pc level. As in System 1 we see a modest colour gradient in the sub-clumps, with those further from the caustic lines appearing bluer; however, in this case the effect is qualitatively weaker. Looking at the resolved CIII] emission we again detect no change in line strength or width throughout the galaxy. 

We then move to the system 10 source, noting that the reconstruction is limited to the Ly$\alpha$ gas component alone, since its continuum does not appear in the \emph{HST} data (likely falling below the detection limit achievable with SNAPshot exposure times).  The galaxy's source-plane appearance is largely similar to that of its image-plane counterpart, though this is partially due to the lower resolution of the MUSE data. In the source plane the Ly$\alpha$ flux appears nearly round and is largely compact, extending to only 1.09 kpc in diameter. Similar to System 2, the galaxy also partially overlaps the caustic curves, resulting in an image-plane merger. However, in this case the overlap only occurs along the tangential caustic, leaving the two images lying closer the the radial critical curve as fully separate objects.  Due to its size, the System 10 source experiences a nearly equal magnification boost over its entire extent, with an average magnification of $\mu = 120$ over the brightest point of the merging pair and a more moderate $\mu = 20$ for the separated images 10.3 and 10.4. Focusing on image 10.4, we measure a line flux density of $1.58 \times 10^{-19}$ erg/s/cm$^2$/\AA\ in a 1\arcsec\ diameter aperture (fully enclosing the observed emission), resulting in an intrinsic line flux of $7.85 \times 10^{-20}$ erg/s/cm$^2$/\AA\ when accounting for the magnification boost. Converting the flux to a luminosity yields a value of $\log(L) = 40.51$, placing the galaxy at the faint edge of the observed luminosity function of similar MUSE-detected $z = 6$ Ly$\alpha$ emitters \citep[e.g][]{deLaVieuville2019,claeyssens2022}.

\section{Dark Matter Science}
\label{sec:DMscience}
While the high magnification experienced by H-U systems makes them ideal targets to study resolved properties of source galaxies, it also turns the galaxies into useful tools for studying aspects of dark matter. In this section, we present our initial investigations into the nature of DM, and discuss how these efforts can be improved in the future, especially with a larger sample of H-U objects.

\subsection{Measuring the central mass slope}
Physical properties of the DM particle (such as its mass or interaction cross-section) can alter the spatial gradient of the distribution of DM. In particular, if DM particles were more relativistic in the early Universe (``warm''; WDM) or have a non-negligible cross-section for self-interactions (SIDM), the central slope will flatten \citep[e.g.][]{maccio2013,ludlow2016,robertson2019} compared to clusters in the standard $\Lambda$CDM model, which DM-only simulations predict to have near-universal NFW \citep{nfw1997} profiles. This is especially true in the innermost ($r < 50$ kpc) regions of galaxy clusters, where DM concentrations are highest. With several lensing constraints probing this crucial inner region, RXJ0437 makes an excellent test case for studying this feature. 

To place our direct slope measurements (Fig.~\ref{fig:profiles}) in a more general context, we compare the value to an NFW model. Specifically, we modify our lens model parametrization scheme (Section~\ref{sec:model}) to include the generalized form of the profile (gNFW) given by: 
\begin{equation}
    \rho(r) = \frac{\rho_s}{(r/r_s)^{\beta} (1 + r/r_s)}
\end{equation}
where $\rho$ is the 3D density of the distribution, $\beta$ the slope of the inner density profile, $r_s$ the characteristic scale radius marking the extent of the inner profile,  and $\rho_s$ the density at the scale radius. By using a gNFW model, we can therefore quantify how the result compares to the standard NFW slope (in which $\beta = 1$). 

As a first step, we construct a new, simplified \textsc{Lenstool} model that removes all previous mass components (the cluster-scale halo, dark clump, BCG, cluster member potentials, and external shear; see Table \ref{tbl:ModParams}) and replaces them with a single, large-scale gNFW halo. By doing so, we fit the entire mass distribution with the gNFW component, allowing us to directly quantify the slope of the total (i.e., DM + baryon) mass profile. Optimizing this model using the hi-res version of the lensing constraints, we find a best-fit inner slope of $\beta = 0.95^{+0.02}_{-0.04}$, which is consistent with a standard NFW profile, and a best-fit scale radius of $r_s = 235^{+39}_{-15}$ kpc, roughly twice the maximum radius of the lensing-constrained region. We present the final model parameters in Table \ref{tbl:ModParamsNFW}. Comparing our results to other systems, we find that the RXJ0437 parameters fall within the 1$\sigma$ uncertainty limits of Abell 1703 (the other known H-U cluster; \citealt{limousin2008}) and within 2$\sigma$ of the mean slope of a set of relaxed clusters with prominent central arcs \citet{newman2013a}. That the total mass profile is consistent with an NFW is not surprising, as \citet{newman2013} find that interactions between the (peakier) baryon-dominated BCG and the (flatter) DM-dominated cluster halo can lead to a standard NFW profile, acting as an up-scaled analogue of the ``bulge-halo conspiracy'' of individual massive elliptical galaxies \citep[e.g.][]{koopmans2006,gavazzi2007}.

We next attempt to uncouple the DM and baryon profiles from one another, using a technique adapted from \citet{limousin2008}. In that work, the authors construct a modified \textsc{Lenstool} model using a fixed PIEMD component to represent the BCG stellar mass (scaling the mass based on the BCG luminosity), and a gNFW profile to represent the DM halo. Here, we also use gNFW and PIEMD components for the DM halo and BCG respectively, but we instead allow the BCG $\sigma$ and $r_{\rm cut}$ parameters to vary (using priors informed by the observed galaxy size, luminosity, and kinematics), to account for the possible existence of extended galaxy light that falls below the \emph{HST} detection limit. After optimizing this new model (Table \ref{tbl:ModParamsNFW}), we find that the gNFW slope flattens considerably, to $\beta = 0.78^{+0.02}_{-0.01}$, while the scale radius is reduced to $r_s = 182^{+4}_{-2}$ kpc, closer to the edge of the strong-lensing region. In this case, the model slope is lower than the \citet{limousin2008} result, who find $\beta \sim 1$ for the DM component, even after separating the BCG mass. However, in that work the authors mention several additional analyses that could ultimately shift their slope value lower. Conversely, our measured slope is largely consistent with the results of \citet{sand2004} and \citet{newman2013}, who also include BCG kinematics when estimating and disentangling stellar mass.

\begin{table*}
  \centering
  \caption{gNFW Lens Models and Best-Fitting Parameters}
  \label{tbl:ModParamsNFW}
  \begin{tabular}{lcrrrrrrr}
    \hline
    Model Name & Component & $\Delta\alpha^{\rm ~a}$& $\Delta\delta^{\rm ~a}$ & $\varepsilon^{\rm ~b}$ / $\gamma$ & $\theta$  / ~~ $\theta_{\gamma}^{\rm ~c}$ & $\beta$ / $r_{\rm core}$ & $r_s$ /~~ $r_{\rm cut}$ & $c$ / ~~~~~~~ $\sigma_0$\\
    (Fit Statistics)$^{\rm ~d}$ &  & (\arcsec) & (\arcsec) & () ~~/ () & ($\deg$) / ($\deg$) & () / (kpc) & (kpc) / (kpc) & () / (km s$^{-1}$)\\
    \hline
    single-halo & gNFW Halo & $  0.24^{+  0.07}_{ -0.04}$ & $  -0.46^{+  0.07}_{ -0.04}$ & $ 0.25^{+ 0.01}_{-0.01}$ &   $-75.6^{+  0.2}_{ -0.2}$ & $ 0.95^{+0.02}_{ -0.04}$ & $235^{+39}_{ -15}$ & $7.3^{+0.8}_{-0.4}$ \\[3pt]
    rms = 0\farcs70 & & & & & & & &\\[3pt]
    $\chi^2/\nu$ = 1.44 & & & & & & & &\\[3pt] 
    $\log~(\mathcal{L}) = -108.07$ & & & & & & & &\\[3pt]
    $\log~(\mathcal{E}) = -140.37$ & & & & & & & &\\
    \hline
    gNFW + dPIE & gNFW Cluster Halo & $  0.25^{+  0.05}_{ -0.05}$ & $  0.64^{+  0.06}_{ -0.08}$ & $ 0.27^{+ 0.01}_{-0.01}$ &   $-72.6^{+  0.6}_{ -0.3}$ & $ 0.78^{+  0.02}_{ -0.01}$ & $182^{+ 4}_{ -2}$ & $9.0^{+0.1}_{-0.2}$ \\[3pt]
    rms = 0\farcs39 & BCG & $[ 0.00]^{\rm ~e}$ & $[  0.00]$ & $[0.37]$ & $[-70.0]$ & $[  0.15]$ & $ 10.6^{+  2.5}_{ -0.1}$ & $153^{+6}_{-1}$ \\[3pt]
    $\chi^2/\nu$ = 1.31 & Dark Clump & $ -0.20^{+  0.05}_{ -0.45}$ & $-12.88^{+  0.98}_{ -0.77}$ & $ 0.89^{+ 0.01}_{-0.04}$ & $-41.1^{+ 3.8}_{-3.7}$ &  $30.0^{+8.3}_{-5.0}$ & $[800.0]$ & $179^{+20}_{-21}$ \\[3pt]
    $\log~(\mathcal{L}) = -22.21$ & L$^{*}$ galaxy &  & & & & $[0.10]$  & $ 13.3^{+ 2.8}_{ -1.8}$ & $175^{+10}_{-9}$\\[3pt]
    $\log~(\mathcal{E}) = -80.52$ & Ext.\ Shear & & & $  0.04^{+  0.01}_{ -0.01}$& $  -15.6^{+  6.7}_{ -5.5}$ & & & \\
\hline
\end{tabular}
\medskip\\
Note 1: For explanations of lettered footnotes, see Table \ref{tbl:ModParams}~~~~~~~~~~~~~~~~~~~~~~~~~~~~~~~~~~~~~~~~~~~~~~~~~~~~~~~~~~~~~~~~~~~~~~~~~~~~~~~~~~~~~~~~~~~~~~~~~~~~~~~~~~~~~~~~~~~~~~~~~~~~~~~~~~~~~~~~~~~~~~~~~~~~~~~~~~~~~~~~~~~~~~~~~\\[1pt]
Note 2: in the final three columns, gNFW components are described by $\beta$, $r_s$, and c (the central mass density slope, scale radius, and concentration parameter), while all other mass components are instead described by the usual $r_{\rm core}$, $r_{\rm cut}$, and $\sigma_0$ parameters~~~~~~~~~~~~~~~~~~~~~~~~~~~~~~~~~~~~~~~~~~~~~~~~~~~~~~~~~~~~~~~~~~~~~~~~~~~~~~~~~~~~~~~~~~~~~~~~
\end{table*}

\subsection{Detecting small-scale substructure}

The nature of the DM particle also influences structure formation, with WDM, SIDM or fuzzy dark matter removing the low mass ($M \leq 10^9 M_\odot$) substructures predicted to be ubiquitous by $\Lambda$CDM \citep{Green2005, Diemand2008, Bose2017, Wang2020}. Primordial halos of mass $\lesssim 10^8\,M_{\odot}$ are too small to have hosted their own star formation \citep{Benitez-Llambay2020}, so would be dark. Searches for them are underway, as perturbations to galaxy-scale lenses \citep{Vegetti2010, Vegetti2012, Hezaveh2016, Ritondale2019, He2022, Nightingale2022}. Cluster lenses a generally less useful, because the complexity of their mass distributions and the wide-separation of their multiple-images make it difficult to identify anomalies such as flux-ratio discrepancies or astrometric offsets. However, H-U configurations are more compact than typical cluster images, and even have an advantage over galaxy-scale lenses:\ the images are well offset from the BCG, thus avoiding contamination from lens light, one of the main systematics in galaxy-galaxy modeling \citep{Pearson2021, Etherington2022}. 

To test the feasibility of detecting DM substructures using H-U systems, we compare flux reconstructions of System 1, using both the best-fit mass model and other models that include simulated point-source substructures.  Specifically, we first project the observed continuum flux of one image (image 1.2, in the upper right corner) to the source plane, then use the result to reconstruct the lens-plane light over the entire H-U region. Measuring residuals between the best-fit model and substructure-based alternatives, we calculate how the point masses alter the light map.  We note that this is a simplified approach to the problem:\ in a full test, we would compare all model reconstructions to the observed data frame, using the substructures as a way to minimize any apparent discrepancies in the initial best-fit reconstruction \citep{He2022, Nightingale2022}.  However, in this preliminary proof-of-concept test, we simply compare model reconstructions directly; this allows us to develop an intuition into how different mass values can affect the results. When comparing models, we classify ``detections'' to be any residual that deviates by  more than $3\sigma$ from the mean sky value, as this would be considered a significant measurement in a real observation. At the present imaging depth, we find that we are only sensitive to substructures with $M \geq 10^9 M_\odot$; this is due largely to the fact that the System 1 continuum is compact and highly structured, increasing the contrast of small perturbations. This is just at the upper mass limit where $\Lambda$CDM models begin to deviate from alternatives, but it is nonetheless useful as a starting point; quadrupling exposure time of galaxy-scale lenses multiplies the number of detectable perturbers by $\sim$2 \citep{Gilman2021, Despali2022}.

\begin{figure*}
    \centering
    \includegraphics[width=\textwidth]{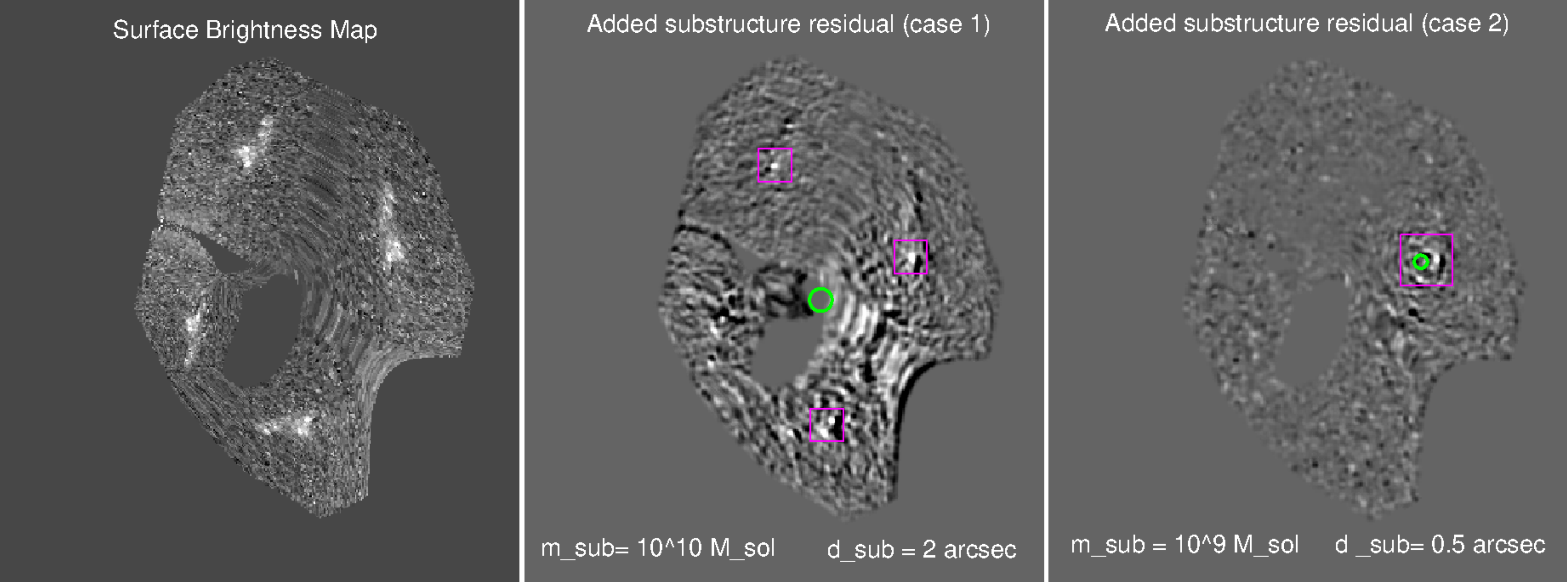}
    \caption{Example of our simulated test for detecting small-scale substructure masses with H-U systems. Left:\ a model reconstruction of the surface brightness of System 1. In this case the observed \emph{HST} flux of one image (Image 2.1, in the top right) is sent to the source plane using the best-fit hi-res mass model; the recreated source is then re-projected back over the entire lens plane, creating an ``idealized'' version of the lens flux. Middle:\ Residual of the left panel and a second model reconstruction, made by adding a moderate ($10^{10} M_{\odot}$) simulated point-source mass to the best-fit model, at the location of the green circle. Areas near the multiple images where the residual deviates by more than $3\sigma$ from the mean sky noise are flagged as detected anomalies (magenta boxes). Right:\ A second residual, using a new model where the substructure is now smaller ($10^9 M_{\odot}$) and placed closer to one image. While a real substructure search will be more involved than this simple test, it nonetheless gives us intuition as to how substructures may affect the observed flux.}
    \label{fig:substructureBlobs}
\end{figure*}

We present examples of our simulation tests in Fig.~\ref{fig:substructureBlobs}, showcasing two configurations that rise just above our detection threshold: one in which a larger mass ($M = 10^{10} M_\odot$) is placed at a moderate (and nearly equal) distance from all lensed images, and another where a smaller mass ($M = 10^9 M_\odot$) lies close to one of them. The effects on image 1.2 are similar (the brightest of the southern clumps is slightly shifted and magnified) and, in the high mass case, two more of the H-U images are also perturbed. Thus, by combining information from each image (which behave as independent probes), the whole of the 2D H-U surface can be used to detect low-mass perturbations \citep[c.f.][]{Chatterjee2018, Rivero2018, Cyr-Racine2019, He2020}. 

\section{Conclusions}
\label{sec:conclusions}

We have investigated strong gravitational lensing by the cluster RX J0437.1+0043 (RXJ0437) using a robust combination of imaging and spectroscopy. Taking advantage of several rare, ``exotic'', Hyperbolic-Umbilic (H-U) gravitational lens systems identified in the field, we infer properties of both the cluster and the highly magnified background galaxies, and use these to explore aspects of the nature of dark matter (DM). Our main results are as follows:

\begin{itemize}
    \item Using a wide mosaic of MUSE data, as well as supporting NIR spectroscopy from Keck/MOSFIRE observations, we measure redshifts for 180 unique targets in the field, consisting of 16 foreground objects, 73 cluster members, and 91 background galaxies -- including 13 multiply imaged systems. Taking the 64 highest-confidence cluster members, we measure a fiducial cluster redshift of $z = 0.2847$ and a line-of-sight cluster velocity dispersion of $1570^{+120}_{-160}$ km s$^{-1}$.

    \item Combining the MUSE data with multi-band DECam and \emph{HST} imaging, we construct a lens model to map the cluster's mass distribution. Thanks to the high density of multiple-image systems, the model is tightly constrained in the $20 \leq r \leq 130$ kpc range, with the H-U images providing critical information about the innermost ($r < 50$ kpc) cluster regions. After optimizing the parameters, our best-fit model is excellent, with an rms error of 0\farcs31. 
    
    \item Generating a mass density map, we see that the structure is dominated by a single cluster-scale DM halo and the central BCG, though we find that adding a second, smaller DM component (a ``dark clump'') in the south, along with a modest level of external shear ($\gamma \sim 0.05$), helps to better fit the southern multiple-image constraints. Contours of the map appear moderately elliptical ($\varepsilon \sim 0.4$) out to large radii ($r \sim 700$ kpc) driven by the closely aligned orientations of the BCG and large-scale DM halo. 
    
    \item Radially averaging the map we construct a surface mass density profile for the cluster, again extending to $\sim700$ kpc in physical space. We measure logarithmic slopes of $\Delta = -0.32$ for $r < 20$ kpc, $\Delta = -0.71$ for ($20 \leq r \leq 130$) kpc; the region containing all multiple image constraints, and $\Delta = -1.68$ for $r > 130$ kpc. Integrating over the complete profile, we measure an aperture mass of ($4.37 \pm 0.43$) $\times 10^{14} M_{\odot}$. We also note that the size of the innermost region (unconstrained by lensing) is smaller in RXJ0437 compared to other clusters of similar mass. This is again due to the presence of the H-U systems. 

    \item Taking advantage of the extreme magnification of H-U systems, we investigate physical properties of their source galaxies. Each galaxy has at least one strong emission line that indicates it is young and undergoing a period of significant star formation. The most prominent object (System 1; $z = 2.9732$) has several high-ionization emission features, including a broad, double-peaked Lyman-$\alpha$ (Ly$\alpha$) line. The Ly$\alpha$ region is considerably extended compared to the continuum, and its two flux peaks are clearly offset from each other and from the stellar component. Taking advantage of the spatial information provided by MOSFIRE/NIR detections of [OIII] emission in two almost perpendicular slits, we conclude that this offset is due to the relative (radial) velocities of subsystems 101 and 102/103.

    \item Focusing on continuum regions, we see that the H-U galaxies appear complex in \emph{HST} imaging, with each showing a considerable amount of substructure.  Reconstructing the galaxies in the source plane we find that they are compact:\ the largest (System 1) is only $\sim$3 kpc in diameter, while the observed substructures are $\sim$200 pc each. Because the magnification in the H-U region is nearly circular (since the sources lie close to both radial and tangential caustic curves) we easily identify substructures at high resolution over the entire 2D galaxy surface, in contrast to most lensed galaxies which are instead enhanced along a single ``preferred'' axis.

    \item Finally, using our cluster mass and source galaxy measurements as a guide, we demonstrate two ways in which clusters containing H-U systems can probe the properties of dark matter:\ 
    
    1.) Separating the total mass profile into luminous and dark components (using the luminosity and flux of the BCG), we measure the central slope of the DM-only distribution using a generalized NFW profile. Our best-fit parameter ($\beta = 0.78^{+0.02}_{-0.01}$) shows considerable flattening compared to the nominal NFW profile predicted by DM-only numerical simulations ($\beta = 1$). However, when we fit the dark and luminous model components together, the \emph{total} density profile ($\beta = 0.95^{+0.02}_{-0.04}$) is consistent with a $\Lambda$CDM universe. We note that such a precision measurement is possible thanks to the extremely tight constraints at small radii provided by the H-U galaxies.

    2.) Using the complex structure of the H-U source galaxies, we demonstrate a proof-of-concept method for detecting small-scale DM substructures in the cluster halo identifying astrometric and flux ratio anomalies. In this way we can determine the substructure mass fraction, another property that is sensitive to the DM model. While we are only sensitive to $\sim10^9 M_{\odot}$ masses at the current imaging depths, deeper data (and additional H-U systems) will make this technique competitive with existing efforts focusing on galaxy-galaxy lenses.     
\end{itemize}

Overall, our work highlights the significant and diverse scientific value of H-U systems, especially in the domains of mass mapping and resolved galaxy properties. In addition, we find  exciting new prospects for studying the nature of dark matter, which is becoming increasingly important in the era of precision cosmology. However, while the results presented here are promising, the total number of known H-U galaxies is still small, and information derived from them may be sensitive to observational or population biases. With only two identified clusters housing H-U systems (RXJ0437 and Abell 1703), it is difficult to know if they are representative of a larger parent distribution. To perform a truly robust statistical analysis with H-U lenses, we will thus need to increase the sample size. To this end, we have already begun searching archival MUSE data for additional cases -- focusing on  Kaleidoscope clusters and the MUSE GTO lensing-cluster atlas \citep{richard2021}. Although this search is still in the early stages, we have already identified $\sim 10$ additional H-U candidates, which will be the subject of a forthcoming paper. 

Improving on the impossibly rare ``once-in-a-sky-survey'' predictions of the past \citep[e.g.~][]{orban2009}, our current and ongoing work suggests that H-U galaxies may be slightly more common, in line with more recent predictions presented in \citet{meena2021a} and \citet{meena2021b}. In order to better understand under which conditions H-U systems are most readily generated, our ongoing work in this area also catalogues the physical properties of the host clusters. This information will be especially beneficial in the near future, when upcoming large surveys such as the \emph{Euclid} Wide Survey \citep{laureijs2011} and the Legacy Survey of Space and Time \citep{lsst2009} will identify thousands of new lensing clusters. 

\section*{Acknowledgements}
DJL, ACE and RM are supported by STFC grants ST/T000244/1 and ST/W002612/1.
MJ and DJL are supported by the United Kingdom Research and Innovation (UKRI) Future Leaders Fellowship `Using Cosmic Beasts to uncover the Nature of Dark Matter' (grant number MR/S017216/1). GM received funding from the European Union’s Horizon 2020 research and innovation programme under the Marie Skłodowska-Curie grant agreement No MARACAS - DLV-896778.

\section*{Data Availability}

The data underlying this article are available in the article itself and its online supplementary material.



\bibliographystyle{mnras}
\bibliography{manuscript} 

\begin{thebibliography}{}
\makeatletter
\relax
\def\mn@urlcharsother{\let\do\@makeother \do\$\do\&\do\#\do\^\do\_\do\%\do\~}
\def\mn@doi{\begingroup\mn@urlcharsother \@ifnextchar [ {\mn@doi@}
  {\mn@doi@[]}}
\def\mn@doi@[#1]#2{\def\@tempa{#1}\ifx\@tempa\@empty \href
  {http://dx.doi.org/#2} {doi:#2}\else \href {http://dx.doi.org/#2} {#1}\fi
  \endgroup}
\def\mn@eprint#1#2{\mn@eprint@#1:#2::\@nil}
\def\mn@eprint@arXiv#1{\href {http://arxiv.org/abs/#1} {{\tt arXiv:#1}}}
\def\mn@eprint@dblp#1{\href {http://dblp.uni-trier.de/rec/bibtex/#1.xml}
  {dblp:#1}}
\def\mn@eprint@#1:#2:#3:#4\@nil{\def\@tempa {#1}\def\@tempb {#2}\def\@tempc
  {#3}\ifx \@tempc \@empty \let \@tempc \@tempb \let \@tempb \@tempa \fi \ifx
  \@tempb \@empty \def\@tempb {arXiv}\fi \@ifundefined
  {mn@eprint@\@tempb}{\@tempb:\@tempc}{\expandafter \expandafter \csname
  mn@eprint@\@tempb\endcsname \expandafter{\@tempc}}}

\bibitem[\protect\citeauthoryear{{Acebron} et~al.,}{{Acebron}
  et~al.}{2022}]{acebron2022}
{Acebron} A.,  et~al., 2022, \mn@doi [\apj] {10.3847/1538-4357/ac3d35}, \href
  {https://ui.adsabs.harvard.edu/abs/2022ApJ...926...86A} {926, 86}

\bibitem[\protect\citeauthoryear{{Andrade}, {Minor}, {Nierenberg}  \&
  {Kaplinghat}}{{Andrade} et~al.}{2019}]{andrade2019}
{Andrade} K.~E.,  {Minor} Q.,  {Nierenberg} A.,   {Kaplinghat} M.,  2019,
  \mn@doi [\mnras] {10.1093/mnras/stz1360}, \href
  {https://ui.adsabs.harvard.edu/abs/2019MNRAS.487.1905A} {487, 1905}

\bibitem[\protect\citeauthoryear{{Bacon} et~al.,}{{Bacon}
  et~al.}{2010}]{bacon2010}
{Bacon} R.,  et~al., 2010, in {McLean} I.~S.,  {Ramsay} S.~K.,   {Takami} H.,
  eds,  Society of Photo-Optical Instrumentation Engineers (SPIE) Conference
  Series Vol. 7735, Ground-based and Airborne Instrumentation for Astronomy
  III. p. 773508, \mn@doi{10.1117/12.856027}

\bibitem[\protect\citeauthoryear{{Bacon}, {Piqueras}, {Conseil}, {Richard}  \&
  {Shepherd}}{{Bacon} et~al.}{2016}]{bacon2016}
{Bacon} R.,  {Piqueras} L.,  {Conseil} S.,  {Richard} J.,   {Shepherd} M.,
  2016, {MPDAF: MUSE Python Data Analysis Framework}, Astrophysics Source Code
  Library, record ascl:1611.003 (\mn@eprint {ascl} {1611.003})

\bibitem[\protect\citeauthoryear{{Beers}, {Flynn}  \& {Gebhardt}}{{Beers}
  et~al.}{1990}]{beers1990}
{Beers} T.~C.,  {Flynn} K.,   {Gebhardt} K.,  1990, \mn@doi [\aj]
  {10.1086/115487}, \href
  {https://ui.adsabs.harvard.edu/abs/1990AJ....100...32B} {100, 32}

\bibitem[\protect\citeauthoryear{{Benitez} \& {Frenk}}{{Benitez} \&
  {Frenk}}{2020}]{Benitez-Llambay2020}
{Benitez} A.,  {Frenk} C.,  2020, \mn@doi [\mnras] {10.1093/mnras/staa2698},
  \href {https://ui.adsabs.harvard.edu/abs/2020MNRAS.498.4887B} {498, 4887}

\bibitem[\protect\citeauthoryear{{Bertin} \& {Arnouts}}{{Bertin} \&
  {Arnouts}}{1996}]{bertin1996}
{Bertin} E.,  {Arnouts} S.,  1996, \mn@doi [\aaps] {10.1051/aas:1996164}, \href
  {https://ui.adsabs.harvard.edu/abs/1996A&AS..117..393B} {117, 393}

\bibitem[\protect\citeauthoryear{{Bose} et~al.,}{{Bose}
  et~al.}{2017}]{Bose2017}
{Bose} S.,  et~al., 2017, \mn@doi [\mnras] {10.1093/mnras/stw2686}, \href
  {https://ui.adsabs.harvard.edu/abs/2017MNRAS.464.4520B} {464, 4520}

\bibitem[\protect\citeauthoryear{{Chatterjee} \& {Koopmans}}{{Chatterjee} \&
  {Koopmans}}{2018}]{Chatterjee2018}
{Chatterjee} S.,  {Koopmans} L.,  2018, \mn@doi [\mnras]
  {10.1093/mnras/stx2674}, \href
  {https://ui.adsabs.harvard.edu/abs/2018MNRAS.474.1762C} {474, 1762}

\bibitem[\protect\citeauthoryear{{Claeyssens} et~al.,}{{Claeyssens}
  et~al.}{2022}]{claeyssens2022}
{Claeyssens} A.,  et~al., 2022, \mn@doi [\aap] {10.1051/0004-6361/202142320},
  \href {https://ui.adsabs.harvard.edu/abs/2022A&A...666A..78C} {666, A78}

\bibitem[\protect\citeauthoryear{{Clowe}, {Brada{\v{c}}}, {Gonzalez},
  {Markevitch}, {Randall}, {Jones}  \& {Zaritsky}}{{Clowe}
  et~al.}{2006}]{clowe2006}
{Clowe} D.,  {Brada{\v{c}}} M.,  {Gonzalez} A.~H.,  {Markevitch} M.,  {Randall}
  S.~W.,  {Jones} C.,   {Zaritsky} D.,  2006, \mn@doi [\apjl] {10.1086/508162},
  \href {https://ui.adsabs.harvard.edu/abs/2006ApJ...648L.109C} {648, L109}

\bibitem[\protect\citeauthoryear{{Cyr-Racine}, {Keeton}  \&
  {Moustakas}}{{Cyr-Racine} et~al.}{2019}]{Cyr-Racine2019}
{Cyr-Racine} F.-Y.,  {Keeton} C.~R.,   {Moustakas} L.~A.,  2019, \mn@doi [PRD]
  {10.1103/PhysRevD.100.023013}, \href
  {https://ui.adsabs.harvard.edu/abs/2019PhRvD.100b3013C} {100, 023013}

\bibitem[\protect\citeauthoryear{{Despali}, {Vegetti}, {White}, {Powell},
  {Stacey}, {Fassnacht}, {Rizzo}  \& {Enzi}}{{Despali}
  et~al.}{2022}]{Despali2022}
{Despali} G.,  {Vegetti} S.,  {White} S. D.~M.,  {Powell} D.~M.,  {Stacey}
  H.~R.,  {Fassnacht} C.~D.,  {Rizzo} F.,   {Enzi} W.,  2022, \mn@doi [\mnras]
  {10.1093/mnras/stab3537}, \href
  {https://ui.adsabs.harvard.edu/abs/2022MNRAS.510.2480D} {510, 2480}

\bibitem[\protect\citeauthoryear{{D{\'\i}az Rivero}, {Dvorkin}, {Cyr-Racine},
  {Zavala}  \& {Vogelsberger}}{{D{\'\i}az Rivero} et~al.}{2018}]{Rivero2018}
{D{\'\i}az Rivero} A.,  {Dvorkin} C.,  {Cyr-Racine} F.-Y.,  {Zavala} J.,
  {Vogelsberger} M.,  2018, \mn@doi [PRD] {10.1103/PhysRevD.98.103517}, \href
  {https://ui.adsabs.harvard.edu/abs/2018PhRvD..98j3517D} {98, 103517}

\bibitem[\protect\citeauthoryear{{Diemand}, {Kuhlen}, {Madau}, {Zemp}, {Moore},
  {Potter}  \& {Stadel}}{{Diemand} et~al.}{2008}]{Diemand2008}
{Diemand} J.,  {Kuhlen} M.,  {Madau} P.,  {Zemp} M.,  {Moore} B.,  {Potter} D.,
    {Stadel} J.,  2008, \mn@doi [\nat] {10.1038/nature07153}, \href
  {https://ui.adsabs.harvard.edu/abs/2008Natur.454..735D} {454, 735}

\bibitem[\protect\citeauthoryear{{Ebeling}, {Edge}, {Allen}, {Crawford},
  {Fabian}  \& {Huchra}}{{Ebeling} et~al.}{2000}]{ebeling2000}
{Ebeling} H.,  {Edge} A.~C.,  {Allen} S.~W.,  {Crawford} C.~S.,  {Fabian}
  A.~C.,   {Huchra} J.~P.,  2000, \mn@doi [\mnras]
  {10.1046/j.1365-8711.2000.03549.x}, \href
  {https://ui.adsabs.harvard.edu/abs/2000MNRAS.318..333E} {318, 333}

\bibitem[\protect\citeauthoryear{{El{\'\i}asd{\'o}ttir}
  et~al.,}{{El{\'\i}asd{\'o}ttir} et~al.}{2007}]{eliasdottir2007}
{El{\'\i}asd{\'o}ttir} {\'A}.,  et~al., 2007, arXiv e-prints, \href
  {https://ui.adsabs.harvard.edu/abs/2007arXiv0710.5636E} {p. arXiv:0710.5636}

\bibitem[\protect\citeauthoryear{{Erb}, {Steidel}  \& {Chen}}{{Erb}
  et~al.}{2018}]{erb2018}
{Erb} D.~K.,  {Steidel} C.~C.,   {Chen} Y.,  2018, \mn@doi [\apjl]
  {10.3847/2041-8213/aacff6}, \href
  {https://ui.adsabs.harvard.edu/abs/2018ApJ...862L..10E} {862, L10}

\bibitem[\protect\citeauthoryear{{Etherington} et~al.,}{{Etherington}
  et~al.}{2022}]{Etherington2022}
{Etherington} A.,  et~al., 2022, \mn@doi [\mnras] {10.1093/mnras/stac2639},
  \href {https://ui.adsabs.harvard.edu/abs/2022MNRAS.517.3275E} {517, 3275}

\bibitem[\protect\citeauthoryear{{Faber} \& {Jackson}}{{Faber} \&
  {Jackson}}{1976}]{faber1976}
{Faber} S.~M.,  {Jackson} R.~E.,  1976, \mn@doi [\apj] {10.1086/154215}, \href
  {https://ui.adsabs.harvard.edu/abs/1976ApJ...204..668F} {204, 668}

\bibitem[\protect\citeauthoryear{{Ferragamo}, {Rubi{\~n}o-Mart{\'\i}n},
  {Betancort-Rijo}, {Munari}, {Sartoris}  \& {Barrena}}{{Ferragamo}
  et~al.}{2020}]{ferragamo2020}
{Ferragamo} A.,  {Rubi{\~n}o-Mart{\'\i}n} J.~A.,  {Betancort-Rijo} J.,
  {Munari} E.,  {Sartoris} B.,   {Barrena} R.,  2020, \mn@doi [\aap]
  {10.1051/0004-6361/201834837}, \href
  {https://ui.adsabs.harvard.edu/abs/2020A&A...641A..41F} {641, A41}

\bibitem[\protect\citeauthoryear{{Ford} et~al.,}{{Ford}
  et~al.}{2003}]{ford2003}
{Ford} H.~C.,  et~al., 2003, in {Blades} J.~C.,  {Siegmund} O. H.~W.,  eds,
  Society of Photo-Optical Instrumentation Engineers (SPIE) Conference Series
  Vol. 4854, Future EUV/UV and Visible Space Astrophysics Missions and
  Instrumentation.. pp 81--94, \mn@doi{10.1117/12.460040}

\bibitem[\protect\citeauthoryear{{Gavazzi}, {Treu}, {Rhodes}, {Koopmans},
  {Bolton}, {Burles}, {Massey}  \& {Moustakas}}{{Gavazzi}
  et~al.}{2007}]{gavazzi2007}
{Gavazzi} R.,  {Treu} T.,  {Rhodes} J.~D.,  {Koopmans} L. V.~E.,  {Bolton}
  A.~S.,  {Burles} S.,  {Massey} R.~J.,   {Moustakas} L.~A.,  2007, \mn@doi
  [\apj] {10.1086/519237}, \href
  {https://ui.adsabs.harvard.edu/abs/2007ApJ...667..176G} {667, 176}

\bibitem[\protect\citeauthoryear{{Ghosh} et~al.,}{{Ghosh}
  et~al.}{2021}]{ghosh2021}
{Ghosh} A.,  et~al., 2021, \mn@doi [\mnras] {10.1093/mnras/stab1196}, \href
  {https://ui.adsabs.harvard.edu/abs/2021MNRAS.506.6144G} {506, 6144}

\bibitem[\protect\citeauthoryear{{Gilman}, {Bovy}, {Treu}, {Nierenberg},
  {Birrer}, {Benson}  \& {Sameie}}{{Gilman} et~al.}{2021}]{Gilman2021}
{Gilman} D.,  {Bovy} J.,  {Treu} T.,  {Nierenberg} A.,  {Birrer} S.,  {Benson}
  A.,   {Sameie} O.,  2021, \mn@doi [\mnras] {10.1093/mnras/stab2335}, \href
  {https://ui.adsabs.harvard.edu/abs/2021MNRAS.507.2432G} {507, 2432}

\bibitem[\protect\citeauthoryear{{Gladders} \& {Yee}}{{Gladders} \&
  {Yee}}{2000}]{gladders2000}
{Gladders} M.~D.,  {Yee} H.~K.~C.,  2000, \mn@doi [\aj] {10.1086/301557}, \href
  {https://ui.adsabs.harvard.edu/abs/2000AJ....120.2148G} {120, 2148}

\bibitem[\protect\citeauthoryear{{Green}, {Hofmann}  \& {Schwarz}}{{Green}
  et~al.}{2005}]{Green2005}
{Green} A.,  {Hofmann} S.,   {Schwarz} D.~J.,  2005, \mn@doi [\jcap]
  {10.1088/1475-7516/2005/08/003}, \href
  {https://ui.adsabs.harvard.edu/abs/2005JCAP...08..003G} {2005, 003}

\bibitem[\protect\citeauthoryear{{Grillo} et~al.,}{{Grillo}
  et~al.}{2015}]{grillo2015}
{Grillo} C.,  et~al., 2015, \mn@doi [\apj] {10.1088/0004-637X/800/1/38}, \href
  {https://ui.adsabs.harvard.edu/abs/2015ApJ...800...38G} {800, 38}

\bibitem[\protect\citeauthoryear{{Harvey}, {Robertson}, {Massey}  \&
  {McCarthy}}{{Harvey} et~al.}{2019}]{harvey2019}
{Harvey} D.,  {Robertson} A.,  {Massey} R.,   {McCarthy} I.~G.,  2019, \mn@doi
  [\mnras] {10.1093/mnras/stz1816}, \href
  {https://ui.adsabs.harvard.edu/abs/2019MNRAS.488.1572H} {488, 1572}

\bibitem[\protect\citeauthoryear{{He} et~al.,}{{He} et~al.}{2022a}]{He2020}
{He} Q.,  et~al., 2022a, \mn@doi [\mnras] {10.1093/mnras/stac191}, \href
  {https://ui.adsabs.harvard.edu/abs/2022MNRAS.511.3046H} {511, 3046}

\bibitem[\protect\citeauthoryear{{He} et~al.,}{{He} et~al.}{2022b}]{He2022}
{He} Q.,  et~al., 2022b, \mn@doi [\mnras] {10.1093/mnras/stac759}, \href
  {https://ui.adsabs.harvard.edu/abs/2022MNRAS.512.5862H} {512, 5862}

\bibitem[\protect\citeauthoryear{{Hezaveh} et~al.,}{{Hezaveh}
  et~al.}{2016}]{Hezaveh2016}
{Hezaveh} Y.,  et~al., 2016, \mn@doi [\apj] {10.3847/0004-637X/823/1/37}, \href
  {https://ui.adsabs.harvard.edu/abs/2016ApJ...823...37H} {823, 37}

\bibitem[\protect\citeauthoryear{{Hinton}}{{Hinton}}{2016}]{hinton2016}
{Hinton} S.,  2016, {MARZ: Redshifting Program} (\mn@eprint {ascl} {1605.001})

\bibitem[\protect\citeauthoryear{{Horne}}{{Horne}}{1986}]{horne1986}
{Horne} K.,  1986, \mn@doi [\pasp] {10.1086/131801}, \href
  {https://ui.adsabs.harvard.edu/abs/1986PASP...98..609H} {98, 609}

\bibitem[\protect\citeauthoryear{{Jauzac} et~al.,}{{Jauzac}
  et~al.}{2016}]{jauzac2016}
{Jauzac} M.,  et~al., 2016, \mn@doi [\mnras] {10.1093/mnras/stw2251}, \href
  {https://ui.adsabs.harvard.edu/abs/2016MNRAS.463.3876J} {463, 3876}

\bibitem[\protect\citeauthoryear{{Jullo} \& {Kneib}}{{Jullo} \&
  {Kneib}}{2009}]{jullo2009}
{Jullo} E.,  {Kneib} J.~P.,  2009, \mn@doi [\mnras]
  {10.1111/j.1365-2966.2009.14654.x}, \href
  {https://ui.adsabs.harvard.edu/abs/2009MNRAS.395.1319J} {395, 1319}

\bibitem[\protect\citeauthoryear{{Jullo}, {Kneib}, {Limousin},
  {El{\'\i}asd{\'o}ttir}, {Marshall}  \& {Verdugo}}{{Jullo}
  et~al.}{2007}]{jullo2007}
{Jullo} E.,  {Kneib} J.~P.,  {Limousin} M.,  {El{\'\i}asd{\'o}ttir} {\'A}.,
  {Marshall} P.~J.,   {Verdugo} T.,  2007, \mn@doi [New Journal of Physics]
  {10.1088/1367-2630/9/12/447}, \href
  {https://ui.adsabs.harvard.edu/abs/2007NJPh....9..447J} {9, 447}

\bibitem[\protect\citeauthoryear{{K{\"a}fer}, {Finoguenov}, {Eckert},
  {Sanders}, {Reiprich}  \& {Nandra}}{{K{\"a}fer} et~al.}{2019}]{kafer2019}
{K{\"a}fer} F.,  {Finoguenov} A.,  {Eckert} D.,  {Sanders} J.~S.,  {Reiprich}
  T.~H.,   {Nandra} K.,  2019, \mn@doi [\aap] {10.1051/0004-6361/201935124},
  \href {https://ui.adsabs.harvard.edu/abs/2019A&A...628A..43K} {628, A43}

\bibitem[\protect\citeauthoryear{{Keeton}, {Kochanek}  \& {Seljak}}{{Keeton}
  et~al.}{1997}]{keeton1997}
{Keeton} C.~R.,  {Kochanek} C.~S.,   {Seljak} U.,  1997, \mn@doi [\apj]
  {10.1086/304172}, \href
  {https://ui.adsabs.harvard.edu/abs/1997ApJ...482..604K} {482, 604}

\bibitem[\protect\citeauthoryear{{Kimble}, {MacKenty}, {O'Connell}  \&
  {Townsend}}{{Kimble} et~al.}{2008}]{kimble2008}
{Kimble} R.~A.,  {MacKenty} J.~W.,  {O'Connell} R.~W.,   {Townsend} J.~A.,
  2008, in {Oschmann} Jacobus~M. J.,  {de Graauw} M. W.~M.,   {MacEwen} H.~A.,
  eds,  Society of Photo-Optical Instrumentation Engineers (SPIE) Conference
  Series Vol. 7010, Space Telescopes and Instrumentation 2008: Optical,
  Infrared, and Millimeter. p. 70101E, \mn@doi{10.1117/12.789581}

\bibitem[\protect\citeauthoryear{{Kneib}, {Ellis}, {Smail}, {Couch}  \&
  {Sharples}}{{Kneib} et~al.}{1996}]{kneib1996}
{Kneib} J.~P.,  {Ellis} R.~S.,  {Smail} I.,  {Couch} W.~J.,   {Sharples} R.~M.,
   1996, \mn@doi [\apj] {10.1086/177995}, \href
  {https://ui.adsabs.harvard.edu/abs/1996ApJ...471..643K} {471, 643}

\bibitem[\protect\citeauthoryear{{Koopmans}, {Treu}, {Bolton}, {Burles}  \&
  {Moustakas}}{{Koopmans} et~al.}{2006}]{koopmans2006}
{Koopmans} L. V.~E.,  {Treu} T.,  {Bolton} A.~S.,  {Burles} S.,   {Moustakas}
  L.~A.,  2006, \mn@doi [\apj] {10.1086/505696}, \href
  {https://ui.adsabs.harvard.edu/abs/2006ApJ...649..599K} {649, 599}

\bibitem[\protect\citeauthoryear{{LSST Science Collaboration} et~al.,}{{LSST
  Science Collaboration} et~al.}{2009}]{lsst2009}
{LSST Science Collaboration} et~al., 2009, arXiv e-prints, \href
  {https://ui.adsabs.harvard.edu/abs/2009arXiv0912.0201L} {p. arXiv:0912.0201}

\bibitem[\protect\citeauthoryear{{Lagattuta} et~al.,}{{Lagattuta}
  et~al.}{2017}]{lagattuta2017}
{Lagattuta} D.~J.,  et~al., 2017, \mn@doi [\mnras] {10.1093/mnras/stx1079},
  \href {https://ui.adsabs.harvard.edu/abs/2017MNRAS.469.3946L} {469, 3946}

\bibitem[\protect\citeauthoryear{{Lagattuta} et~al.,}{{Lagattuta}
  et~al.}{2019}]{lagattuta2019}
{Lagattuta} D.~J.,  et~al., 2019, \mn@doi [\mnras] {10.1093/mnras/stz620},
  \href {https://ui.adsabs.harvard.edu/abs/2019MNRAS.485.3738L} {485, 3738}

\bibitem[\protect\citeauthoryear{{Lagattuta} et~al.,}{{Lagattuta}
  et~al.}{2022}]{lagattuta2022}
{Lagattuta} D.~J.,  et~al., 2022, \mn@doi [\mnras] {10.1093/mnras/stac418},
  \href {https://ui.adsabs.harvard.edu/abs/2022MNRAS.514..497L} {514, 497}

\bibitem[\protect\citeauthoryear{{Laureijs} et~al.,}{{Laureijs}
  et~al.}{2011}]{laureijs2011}
{Laureijs} R.,  et~al., 2011, arXiv e-prints, \href
  {https://ui.adsabs.harvard.edu/abs/2011arXiv1110.3193L} {p. arXiv:1110.3193}

\bibitem[\protect\citeauthoryear{{Limousin} et~al.,}{{Limousin}
  et~al.}{2007}]{limousin2007}
{Limousin} M.,  et~al., 2007, \mn@doi [\apj] {10.1086/521293}, \href
  {https://ui.adsabs.harvard.edu/abs/2007ApJ...668..643L} {668, 643}

\bibitem[\protect\citeauthoryear{{Limousin} et~al.,}{{Limousin}
  et~al.}{2008}]{limousin2008}
{Limousin} M.,  et~al., 2008, \mn@doi [\aap] {10.1051/0004-6361:200809646},
  \href {https://ui.adsabs.harvard.edu/abs/2008A&A...489...23L} {489, 23}

\bibitem[\protect\citeauthoryear{{Ludlow} et~al.,}{{Ludlow}
  et~al.}{2017}]{ludlow2016}
{Ludlow} A.~D.,  et~al., 2017, \mn@doi [\prl] {10.1103/PhysRevLett.118.161103},
  \href {https://ui.adsabs.harvard.edu/abs/2017PhRvL.118p1103L} {118, 161103}

\bibitem[\protect\citeauthoryear{{Macci{\`o}}, {Ruchayskiy}, {Boyarsky}  \&
  {Mu{\~n}oz-Cuartas}}{{Macci{\`o}} et~al.}{2013}]{maccio2013}
{Macci{\`o}} A.~V.,  {Ruchayskiy} O.,  {Boyarsky} A.,   {Mu{\~n}oz-Cuartas}
  J.~C.,  2013, \mn@doi [\mnras] {10.1093/mnras/sts078}, \href
  {https://ui.adsabs.harvard.edu/abs/2013MNRAS.428..882M} {428, 882}

\bibitem[\protect\citeauthoryear{{Mahler} et~al.,}{{Mahler}
  et~al.}{2018}]{mahler2018}
{Mahler} G.,  et~al., 2018, \mn@doi [\mnras] {10.1093/mnras/stx1971}, \href
  {https://ui.adsabs.harvard.edu/abs/2018MNRAS.473..663M} {473, 663}

\bibitem[\protect\citeauthoryear{{Mahler} et~al.,}{{Mahler}
  et~al.}{2023}]{Mahler2023}
{Mahler} G.,  et~al., 2023, \mn@doi [\apj] {10.3847/1538-4357/acaea9}, \href
  {https://ui.adsabs.harvard.edu/abs/2023ApJ...945...49M} {945, 49}

\bibitem[\protect\citeauthoryear{{Massey} et~al.,}{{Massey}
  et~al.}{2018}]{massey2018}
{Massey} R.,  et~al., 2018, \mn@doi [\mnras] {10.1093/mnras/sty630}, \href
  {https://ui.adsabs.harvard.edu/abs/2018MNRAS.477..669M} {477, 669}

\bibitem[\protect\citeauthoryear{{Meena} \& {Bagla}}{{Meena} \&
  {Bagla}}{2020}]{meena2020}
{Meena} A.~K.,  {Bagla} J.~S.,  2020, \mn@doi [\mnras] {10.1093/mnras/stz3632},
  \href {https://ui.adsabs.harvard.edu/abs/2020MNRAS.492.3294M} {492, 3294}

\bibitem[\protect\citeauthoryear{{Meena} \& {Bagla}}{{Meena} \&
  {Bagla}}{2021}]{meena2021a}
{Meena} A.~K.,  {Bagla} J.~S.,  2021, \mn@doi [\mnras] {10.1093/mnras/stab577},
  \href {https://ui.adsabs.harvard.edu/abs/2021MNRAS.503.2097M} {503, 2097}

\bibitem[\protect\citeauthoryear{{Meena}, {Ghosh}, {Bagla}  \&
  {Williams}}{{Meena} et~al.}{2021}]{meena2021b}
{Meena} A.~K.,  {Ghosh} A.,  {Bagla} J.~S.,   {Williams} L. L.~R.,  2021,
  \mn@doi [\mnras] {10.1093/mnras/stab1807}, \href
  {https://ui.adsabs.harvard.edu/abs/2021MNRAS.506.1526M} {506, 1526}

\bibitem[\protect\citeauthoryear{{Navarro}, {Frenk}  \& {White}}{{Navarro}
  et~al.}{1997}]{nfw1997}
{Navarro} J.~F.,  {Frenk} C.~S.,   {White} S. D.~M.,  1997, \mn@doi [\apj]
  {10.1086/304888}, \href
  {https://ui.adsabs.harvard.edu/abs/1997ApJ...490..493N} {490, 493}

\bibitem[\protect\citeauthoryear{{Newman}, {Treu}, {Ellis}, {Sand}, {Nipoti},
  {Richard}  \& {Jullo}}{{Newman} et~al.}{2013a}]{newman2013a}
{Newman} A.~B.,  {Treu} T.,  {Ellis} R.~S.,  {Sand} D.~J.,  {Nipoti} C.,
  {Richard} J.,   {Jullo} E.,  2013a, \mn@doi [\apj]
  {10.1088/0004-637X/765/1/24}, \href
  {https://ui.adsabs.harvard.edu/abs/2013ApJ...765...24N} {765, 24}

\bibitem[\protect\citeauthoryear{{Newman}, {Treu}, {Ellis}, {Sand }, {Nipoti},
  {Richard}  \& {Jullo}}{{Newman} et~al.}{2013b}]{newman2013}
{Newman} A.~B.,  {Treu} T.,  {Ellis} R.~S.,  {Sand } D.~J.,  {Nipoti} C.,
  {Richard} J.,   {Jullo} E.,  2013b, \mn@doi [\apj]
  {10.1088/0004-637X/765/1/24}, \href
  {https://ui.adsabs.harvard.edu/abs/2013ApJ...765...24N} {765, 24}

\bibitem[\protect\citeauthoryear{Nightingale et~al.,}{Nightingale
  et~al.}{2022}]{Nightingale2022}
Nightingale J.~W.,  et~al., 2022, Scanning For Dark Matter Subhalos in Hubble
  Space Telescope Imaging of 54 Strong Lenses

\bibitem[\protect\citeauthoryear{{Oke}}{{Oke}}{1974}]{oke74}
{Oke} J.~B.,  1974, \mn@doi [\apjs] {10.1086/190287}, \href
  {https://ui.adsabs.harvard.edu/abs/1974ApJS...27...21O} {27, 21}

\bibitem[\protect\citeauthoryear{{Orban de Xivry} \& {Marshall}}{{Orban de
  Xivry} \& {Marshall}}{2009}]{orban2009}
{Orban de Xivry} G.,  {Marshall} P.,  2009, \mn@doi [\mnras]
  {10.1111/j.1365-2966.2009.14925.x}, \href
  {https://ui.adsabs.harvard.edu/abs/2009MNRAS.399....2O} {399, 2}

\bibitem[\protect\citeauthoryear{{Patr{\'\i}cio} et~al.,}{{Patr{\'\i}cio}
  et~al.}{2016}]{patricio2016}
{Patr{\'\i}cio} V.,  et~al., 2016, \mn@doi [\mnras] {10.1093/mnras/stv2859},
  \href {https://ui.adsabs.harvard.edu/abs/2016MNRAS.456.4191P} {456, 4191}

\bibitem[\protect\citeauthoryear{{Pearson}, {Maresca}, {Li}  \&
  {Dye}}{{Pearson} et~al.}{2021}]{Pearson2021}
{Pearson} J.,  {Maresca} J.,  {Li} N.,   {Dye} S.,  2021, \mn@doi [\mnras]
  {10.1093/mnras/stab1547}, \href
  {https://ui.adsabs.harvard.edu/abs/2021MNRAS.505.4362P} {505, 4362}

\bibitem[\protect\citeauthoryear{{Petters}, {Levine}  \&
  {Wambsganss}}{{Petters} et~al.}{2001}]{petters01}
{Petters} A.~O.,  {Levine} H.,   {Wambsganss} J.,  2001, {Singularity theory
  and gravitational lensing}

\bibitem[\protect\citeauthoryear{{Piqueras}, {Conseil}, {Shepherd}, {Bacon},
  {Leclercq}  \& {Richard}}{{Piqueras} et~al.}{2017}]{piqueras2017}
{Piqueras} L.,  {Conseil} S.,  {Shepherd} M.,  {Bacon} R.,  {Leclercq} F.,
  {Richard} J.,  2017, arXiv e-prints, \href
  {https://ui.adsabs.harvard.edu/abs/2017arXiv171003554P} {p. arXiv:1710.03554}

\bibitem[\protect\citeauthoryear{{Richard} et~al.,}{{Richard}
  et~al.}{2021}]{richard2021}
{Richard} J.,  et~al., 2021, \mn@doi [\aap] {10.1051/0004-6361/202039462},
  \href {https://ui.adsabs.harvard.edu/abs/2021A&A...646A..83R} {646, A83}

\bibitem[\protect\citeauthoryear{{Ritondale}, {Vegetti}, {Despali}, {Auger},
  {Koopmans}  \& {McKean}}{{Ritondale} et~al.}{2019}]{Ritondale2019}
{Ritondale} E.,  {Vegetti} S.,  {Despali} G.,  {Auger} M.~W.,  {Koopmans} L.,
  {McKean} J.~P.,  2019, \mn@doi [\mnras] {10.1093/mnras/stz464}, \href
  {https://ui.adsabs.harvard.edu/abs/2019MNRAS.485.2179R} {485, 2179}

\bibitem[\protect\citeauthoryear{{Robertson}, {Harvey}, {Massey}, {Eke},
  {McCarthy}, {Jauzac}, {Li}  \& {Schaye}}{{Robertson}
  et~al.}{2019}]{robertson2019}
{Robertson} A.,  {Harvey} D.,  {Massey} R.,  {Eke} V.,  {McCarthy} I.~G.,
  {Jauzac} M.,  {Li} B.,   {Schaye} J.,  2019, \mn@doi [\mnras]
  {10.1093/mnras/stz1815}, \href
  {https://ui.adsabs.harvard.edu/abs/2019MNRAS.488.3646R} {488, 3646}

\bibitem[\protect\citeauthoryear{{Sand}, {Treu}, {Smith}  \& {Ellis}}{{Sand}
  et~al.}{2004}]{sand2004}
{Sand} D.~J.,  {Treu} T.,  {Smith} G.~P.,   {Ellis} R.~S.,  2004, \mn@doi
  [\apj] {10.1086/382146}, \href
  {https://ui.adsabs.harvard.edu/abs/2004ApJ...604...88S} {604, 88}

\bibitem[\protect\citeauthoryear{{Schneider}, {Ehlers}  \& {Falco}}{{Schneider}
  et~al.}{1992}]{schneider1992}
{Schneider} P.,  {Ehlers} J.,   {Falco} E.~E.,  1992, {Gravitational Lenses},
  \mn@doi{10.1007/978-3-662-03758-4.
}

\bibitem[\protect\citeauthoryear{{Taylor}, {Massey}, {Jauzac}, {Courbin},
  {Harvey}, {Joseph}  \& {Robertson}}{{Taylor} et~al.}{2017}]{taylor2017}
{Taylor} P.,  {Massey} R.,  {Jauzac} M.,  {Courbin} F.,  {Harvey} D.,  {Joseph}
  R.,   {Robertson} A.,  2017, \mn@doi [\mnras] {10.1093/mnras/stx855}, \href
  {https://ui.adsabs.harvard.edu/abs/2017MNRAS.468.5004T} {468, 5004}

\bibitem[\protect\citeauthoryear{{Vegetti}, {Koopmans}, {Bolton}, {Treu}  \&
  {Gavazzi}}{{Vegetti} et~al.}{2010}]{Vegetti2010}
{Vegetti} S.,  {Koopmans} L.,  {Bolton} A.,  {Treu} T.,   {Gavazzi} R.,  2010,
  \mn@doi [\mnras] {10.1111/j.1365-2966.2010.16865.x}, \href
  {https://ui.adsabs.harvard.edu/abs/2010MNRAS.408.1969V} {408, 1969}

\bibitem[\protect\citeauthoryear{{Vegetti}, {Lagattuta}, {McKean}, {Auger},
  {Fassnacht}  \& {Koopmans}}{{Vegetti} et~al.}{2012}]{Vegetti2012}
{Vegetti} S.,  {Lagattuta} D.~J.,  {McKean} J.~P.,  {Auger} M.~W.,  {Fassnacht}
  C.~D.,   {Koopmans} L.,  2012, \mn@doi [\nat] {10.1038/nature10669}, \href
  {https://ui.adsabs.harvard.edu/abs/2012Natur.481..341V} {481, 341}

\bibitem[\protect\citeauthoryear{{Wang}, {Bose}, {Frenk}, {Gao}, {Jenkins},
  {Springel}  \& {White}}{{Wang} et~al.}{2020}]{Wang2020}
{Wang} J.,  {Bose} S.,  {Frenk} C.~S.,  {Gao} L.,  {Jenkins} A.,  {Springel}
  V.,   {White} S.~D.~M.,  2020, \mn@doi [\nat] {10.1038/s41586-020-2642-9},
  \href {https://ui.adsabs.harvard.edu/abs/2020Natur.585...39W} {585, 39}

\bibitem[\protect\citeauthoryear{{de La Vieuville} et~al.,}{{de La Vieuville}
  et~al.}{2019}]{deLaVieuville2019}
{de La Vieuville} G.,  et~al., 2019, \mn@doi [\aap]
  {10.1051/0004-6361/201834471}, \href
  {https://ui.adsabs.harvard.edu/abs/2019A&A...628A...3D} {628, A3}

\makeatother
\end{thebibliography}




\appendix

\section{Sample Redshift Catalogue}
\label{sec:sampleCat}

In this appendix we present a truncated version of our final redshift catalogue (Table \ref{tbl:zcat}), which is included in the online supplementary material to this manuscript. The format lagely follows the catalogue gresented in \citet{lagattuta2022}, though we will again briefly describe each column here:

\begin{itemize}
    \item \textbf{ID}: a numerical identifier for each object, typically matched to a SExtractor detection run. \smallskip
    
    \item \textbf{Source}: An indication of how the object was detected. \emph{Prior} sources are identified in \emph{HST} images, while \emph{muselet} sources are only found in the MUSE data. (see Section \ref{sec:redshifts})\smallskip
    
    \item \textbf{RA, Dec}: 2D spatial coordinates for each object.\smallskip
    
    \item \textbf{z}: The measured redshift of each object.\smallskip
    
    \item \textbf{z$_{\rm conf}$}: An assessment of the reliability of the redshift measurement, from low ($z_{\rm conf} = 1$) to high ($z_{\rm conf} = 3$). We categorize each confidence level as follows: \smallskip

    Confidence 1: the redshift is based on a single ambiguous or low-SNR emission line, or several low SNR absorption features.\smallskip

    Confidence 2: the redshift is based on a single emission line without additional information, several moderate S/N absorption features, or a Confidence 1 detection whose redshift confidence is increased by the identification of a multiply imaged system.\smallskip

    Confidence 3: the redshift is based on multiple clear spectral features, or on a single high S/N emission line with additional information (e.g., an obvious asymmetry in the line profile or a clear non-detection in {\it HST} bands blueward of the line). \smallskip
    
    \item \textbf{Mult ID}: The unique image number (see Table \ref{tbl:Multi-Images}) given to the object if it is part of a multiply imaged system.
\end{itemize}

\begin{table}
    \centering
    \caption{Sample Redshift Catalogue}
    \label{tbl:zcat}
    \begin{tabular}{l|l|l|l|l|l|l}
    \hline
    ID & Source & RA    &  Dec  & $z$ & $z_{\rm conf}$ & Mult ID \\
       &        & [deg] & [deg] &     &                &         \\
    \hline
        148	& PRIOR	  & 69.278130 & 0.734774 & 0.0000 & 3 & --   \\
        143	& MUSELET & 69.287492 & 0.723762 & 0.2276 & 3 & --   \\
        347	& PRIOR	  & 69.299119 & 0.738344 & 0.2706 & 2 & --   \\
        420	& PRIOR	  & 69.302130 & 0.731618 & 0.2812 & 1 & --   \\
        113	& PRIOR	  & 69.282625 & 0.726220 & 0.8922 & 3 & --   \\
        384	& PRIOR	  & 69.288286 & 0.731356 & 2.9719 & 3 & 1.4  \\
        220	& PRIOR	  & 69.286482 & 0.731973 & 2.9729 & 3 & 1.2  \\
        364	& PRIOR	  & 69.299554 & 0.736776 & 3.7420 & 1 & --   \\
        114	& MUSELET & 69.298123 & 0.742108 & 4.6039 & 2 & --   \\
        243	& MUSELET & 69.288038 & 0.732499 & 6.0152 & 3 & 10.4 \\
        \hline
    \end{tabular}
\end{table}


\bsp	
\label{lastpage}
\end{document}